\documentclass[12pt,a4paper]{JHEP}
\usepackage{amssymb,amsfonts}
\usepackage{latexsym}
\usepackage{amsmath}
\usepackage{fancybox}
\usepackage{frame}
\usepackage{slashed}
\usepackage{ulem}
\usepackage{color}
\usepackage{cite}

\relax
\def\be{\begin{equation}}
\def\ee{\end{equation}}
\def\bea{\begin{eqnarray}}
\def\eea{\end{eqnarray}}

\newcommand{\bear}{\begin{eqnarray}}

\newcommand{\eear}{\end{eqnarray}}

\def\sq
\def\y{\psi}

\DeclareMathOperator{\Tr}{Tr}

\DeclareMathOperator{\diag}{diag}

\DeclareMathOperator{\spn}{span}

\def\bR {\mathbb{R}}

\newcommand{\beq}{\begin{equation}}
\newcommand{\eeq}{\end{equation}}
\newcommand{\bal}{\begin{equation}\begin{aligned}}
\newcommand{\eal}{\end{aligned}\end{equation}}
\newcommand{\half}{\frac{1}{2}}

\newcommand{\eqn}[1]{(\ref{#1})}
\newcommand{\nn}{\nonumber}

\newcommand{\cK}{{\mathcal K}}

\newcommand{\cN}{{\mathcal N}}

\title{\boldmath On Classical Solutions of \\$4d$ Supersymmetric Higher Spin Theory}

\author{Jun Bourdier and Nadav Drukker 
\\
Department of Mathematics, King's College London,\\
The Strand, WC2R 2LS, London, United-Kingdom\\
\email{jun.bourdier@kcl.ac.uk},
\email{nadav.drukker@gmail.com}}

\preprint{} 

\abstract{We present a simple construction of solutions to the supersymmetric 
higher spin theory based on solutions to bosonic theories. We illustrate this for the 
case of the Didenko-Vasiliev solution in arXiv:0906.3898, for which we have 
found a striking simplification where the higher-spin connection takes the 
vacuum value. Studying these solutions 
further, we check under which conditions they preserve some supersymmetry in the bulk, 
and when they are compatible with the boundary conditions conjectured to be 
dual to certain 3d SUSY Chern-Simons-matter theories. 
We perform the analysis for a variety of theories with $\cN=2$, $\cN=3$, $\cN=4$ and $\cN =6$ 
and find a rich spectrum of $1/4$, $1/3$ and $1/2$-BPS solutions.}

\keywords{AdS/CFT, Holography, Higher-Spin, Vector Model, Classical Solutions}

\begin{document}

\section{Introduction}
\label{sec:intro}

Vasiliev's theory of higher-spins \cite{original} consistently describes the interaction 
of fields of any positive spin, and apart from its intrinsic mathematical interest, it was 
first investigated as a possible contender for an unbroken phase of string theory. 
The theory has seen a surge of interest in recent years due to the fact that it can 
consistently be formulated on negatively curved spaces and thus a ripe candidate to 
study the holographic principle 
\cite{Sezgin,Klebanov,threepoint,shenker,holography,holographyreview}.

One of the most appealing features of higher spin holography is that it does not 
require the full machinery of string theory and in particular supersymmetry. The bulk 
side is a theory of higher spins, including the graviton. Of course, there are disadvantages 
as well, with no quantum definition of the theory as of yet. Still, though supersymmetry is 
not required, it is easy to incorporate and allows to holographically study supersymmetric 
boundary theories. It is also useful in order to see how string theory and higher spin theory 
are related.

In particular, a very wide range of supersymmetric higher spin holographic duals were 
proposed in~\cite{triality}. These are four dimensional bulk theories dual to 3d supersymmetric 
Chern-Simons-matter theories. One example even covers a limit of ABJ theory \cite{ABJ}, 
giving a new perspective for studying the relationship of Vasiliev's theory to string theory.

Exact solutions to Vasiliev's higher spin equations in $4$-dimensions are scarce. 
Apart from the vacuum solution, the first exact solution was given by Sezgin and 
Sundell in~\cite{firstexact}. A few years later, Didenko and Vasiliev obtained 
in~\cite{DV} an exact solution with many similarities to an extremal 
black hole. Iazeolla and Sundell then generalized some aspect of the construction and obtained six families of exact solutions in~\cite{families}. 
More recently, a new family of solutions was found by Gubser and Song 
in~\cite{GubserSong}. In this paper we wish to study the classical solutions of 
the supersymmetric higher spin theories.

Classical solutions of interacting theories like gravity and non-abelian gauge theory 
provide important reference points for the understanding of the theories. One can study 
the fluctuations about them and consider them as alternative vacua. In particular 
for supersymmetric theories, solutions which preserve a large fraction of supersymmetry 
may be insensitive to quantum effects and followed from weak to strong coupling, 
like D-branes in string theory. In the holographic setting they can be studied by both a 
weakly coupled gauge theory and a weakly curved gravitational theory.

With this in mind, we study the embedding of classical solutions of bosonic higher spin 
theory into the supersymmetric theory. After presenting some general features of 
such embeddings, we focus on the case of the solution of Didenko and Vasiliev (DV) \cite{DV}. 
By studying the solution in a particular coordinate system and choosing a specific 
time-like Killing vector which appears in the solution, we note a remarkable 
simplification to the form of the solution --- one of the master fields of the theory 
takes the same value as on the $AdS$ vacuum. We also generalized the solution 
to more general bosonic higher spin theories, by allowing a non-zero parity breaking 
phase $\theta_0$.\footnote{The solutions that are found in \cite{families} already take into account the phase $\theta_0$, but are presented in a different gauge.}

This allows us to study such solutions in the supersymmetric higher spin theory, 
and as already noted in~\cite{DV}, preserving supersymmetry requires embedding two different 
(but closely related) bosonic solutions into the supersymmetric theory. In theories 
with extended supersymmetry we find a rich structure of solutions which preserve 
different fractions of the supersymmetries of the bulk $AdS_4$ vacuum.

To define the higher spin theory unambiguously requires choosing boundary conditions for the 
fields of lowest spin, as was advocated in~\cite{ABJ}. Depending on the boundary conditions, 
the same bulk theory is dual to many different $3$d vector Chern-Simons theories with varying 
amount of supersymmetry. We therefore examine the asymptotics of the DV solution and 
its possible embeddings in the proposed holographic duals of certain 3d theories with 
$\cN=2$, $\cN=3$, $\cN=4$ and $\cN = 6$ supersymmetry. We find an intricate structure of families of
$1/4$-BPS, $1/3$-BPS and $1/2$-BPS solutions depending on varying numbers of 
free parameters. This structure of the solutions is quite nicely correlated to the global 
symmetries of the holographic duals, and we comment briefly on how to match them 
to the spectrum of protected operators in the dual theory.

We tried to organize the paper in a natural logical order and included most technical 
details (including the review and simplification of the DV solution) in appendices.

\section{Supersymmetric Higher Spin Theory: a Brief Review}
\label{sec:review}
Our aim here is to give a very short introduction to Vasiliev's theory of higher-spin, 
omitting a lot of subtleties for the sake of clarity. More details can be found in the 
various 
existing 
reviews such as \cite{96review,99review,Bekaert:2010hw}.

\subsection{Master Fields of the Theory}
We present in this section the higher-spin theory following \cite{triality}. 
The theory consistently describes the interactions of an infinite tower of real massless 
fields in $AdS_4$ carrying integer (and in the presence of supersymmetry also half integer) spin
with 
a scalar field of mass $m^2=-2\lambda^{-2}$, where $\lambda$ is the $AdS$ curvature radius. 
The fields are packaged into the 
master fields $(W, B, S)$, which all depend on the following sets of variables
\begin{itemize}
\item $x$: the space-time coordinates.
\item $(y_\alpha ,\bar {y}_{\dot \alpha})$: internal bosonic coordinates. 
Roughly speaking, the coefficients in the expansion of the master fields in 
these coordinates contain the dynamical (or auxiliary) fields. They are collectively denoted $Y$.
\item $(z_\alpha ,\bar {z}_{\dot \alpha})$: similar in nature to the $Y$ coordinates, 
but 
have a very different role as they are introduced to turn on interactions in an 
explicitly gauge-invariant way. They are collectively denoted $Z$.
\item $\vartheta^i$: variables satisfying a Clifford algebra 
$\left\{\vartheta^i, \vartheta^j \right\}=2 \delta^{ij}$. 
Taking $i=1,\cdots,n$ turns every field into a $2^n$ component 
superfield. We recover the bosonic theory by setting $n=0$.
\end{itemize}
The master fields have the following content
\begin{itemize}
\item $W(Y,Z|x,\vartheta)$: is the higher-spin connection, containing the massless 
higher-spin gauge fields of spin $s \geq 1$, as well as auxiliary fields. 
It is a space-time one-form.
\item $B(Y,Z|x,\vartheta)$: contains the curvature of the fields, such as the Weyl 
tensor and its higher-spin generalisation, as well as the massive scalar, massless 
fermion and Maxwell field. It is a space-time zero-form.
\item $S(Y,Z|x,\vartheta)$: is also introduced to turn on interactions, and is purely 
auxiliary. It is a space-time zero-from, but a one-form in $Z$-space
\beq
S=S_\alpha d z^\alpha+\bar S_{\dot \alpha} d \bar{z}^{\dot \alpha}
\eeq
\end{itemize}
One can combine the the two one-forms into a single field
\begin{equation}
\mathcal{A}=W+S\,,
\end{equation}
It is required that $\left\{dx^\mu , d z^\alpha \right\}=\left\{dx^\mu , d \bar{z}^{\dot \alpha} \right\}=0$ 
and we will also define 
$dz^2=\epsilon_{\alpha \beta} dz^\alpha dz^\beta$
and 
$d \bar{z}^2=\epsilon_{\dot \alpha \dot \beta} d\bar{z}^{\dot \alpha} d\bar{z}^{\dot \beta}$.

\subsection{Equations of Motion}

The Vasiliev equations for interacting higher spin fields are
\begin{align}
&\mathcal F \equiv d \mathcal A-\mathcal A \wedge_\star \mathcal A 
=-f_\star \left(B \star v \right) dz^2-\bar{f}_\star \left(B \star \bar{v}\Gamma \right)d \bar{z}^2\,,
\label{masterfull21} \\
&d B-\mathcal{A} \star B+B \star \pi (\mathcal{A})=0\,.
\label{masterfull22}
\end{align}
This deceptively elegant form requires all the following definitions
\begin{itemize}
\item 
Multiplication is performed using
the star product 
(see Appendix~\ref{sec:star})
\end{itemize}
\beq
\label{starproduct}
\Phi(Y,Z) \star \Theta(Y,Z)=\Phi(Y,Z) \exp \left[-\epsilon^{\alpha\beta} 
\big({\mathop\partial^\leftarrow}_{y^\alpha}+{\mathop\partial^\leftarrow}_{z^\alpha} \big) 
\big({\mathop\partial^\rightarrow}_{y^\beta}-{\mathop\partial^\rightarrow}_{z^\beta} \big)
+\text{c.c.} \right] \Theta(Y,Z)\,.
\eeq
\begin{itemize}
\item A function $f(X)=1+X e^{i \theta(X)}$ determining the interactions of the theory. 
A-priori $f$ can be a general function, but it was shown in~\cite{triality,Bosonicdef} that 
after field redefinitions and imposing reality conditions it can be written in this form with 
$\theta$ an even function of $X$ 
(and all products and exponentiation have to be done with the star product, 
hence the subscript $f_*$).

\item The Kleiniens
$v=e^{z_\alpha y^\alpha}$, $\bar{v}=e^{\bar{z}_{\dot \alpha} \bar{y}^{\dot \alpha}}$
which satisfy
\end{itemize}
\bal
v \star v&=1\,,
\qquad&&
v \star \Phi(Y,Z) \star v=\Phi (-\gamma Y,-\gamma Z)\,, \\
\bar{v} \star \bar{v}&=1\,,
&&
\bar{v} \star \Phi (Y,Z) \star \bar{v}=\Phi (\gamma Y,\gamma Z) \, , \\
\qquad
\gamma(\circ_\alpha)&=\circ_\alpha\,,
&\qquad
&\gamma(\bar\circ_{\dot\alpha})=-\bar\circ_{\dot\alpha}
\eal
\begin{itemize}

\item The generalized twist operators $\pi$ and $\bar\pi$ acting by
\bal
\pi \left(\Phi (Y,Z,dZ) \right)&=\Phi (-\gamma Y,-\gamma Z,-\gamma d Z) \label{twist}\,, \\
\bar{\pi} \left(\Phi (Y,Z,dZ) \right)&=\Phi (\gamma Y, \gamma Z,\gamma d Z)
\eal
One should be careful with the fact that 
for 
a 1-form in 
$(z_\alpha, \bar z_{\dot \alpha})$, this generalized twist operator is 
equivalent to conjugation by $v$ \emph{and} the flipping of the sign of $dz^\alpha$.

\item The chirality operator
$\Gamma=i^{\frac{n(n-1)}{2}} \vartheta^1 \vartheta^2 \dots \vartheta^n$.

\end{itemize}

The equations of motion \eqref{masterfull21}, \eqref{masterfull22} are invariant under 
a very large set of gauge transformations
\beq
\label{gauge0}
\delta \mathcal A=d\epsilon-\left[\mathcal A , \epsilon \right]_\star\,,\qquad 
\delta B=\epsilon \star B-B \star \, \pi\left(\epsilon\right)\,.
\eeq
where the gauge parameter $\epsilon(Y,Z|x,\vartheta)$ is a zero-form 
which satisfies the same reality conditions and truncations as $W$.

In components, the equations of motion \eqref{masterfull21}-\eqref{masterfull22} read
\begin{subequations}
\begin{align}
& dW-W \wedge_\star W=0\,, \label{masterfull1} \\
& dB-W \star B+B \star \pi(W)=0\,, \label{masterfull2}\hskip-5mm \\
& dS_\alpha-\left[W,S_\alpha\right]_\star=0\,,
&&d\bar{S}_{\dot \alpha}-\left[W, \bar{S}_{\dot \alpha}\right]_\star=0\,, 
\label{masterfull3} \\
& B \star \pi(S_\alpha)+S_\alpha \star B=0\,,
&&B \star \pi(\bar{S}_{\dot \alpha})-\bar{S}_{\dot \alpha} \star B=0\,, \label{masterfull5} \\
&S_\alpha \star S^\alpha=2 f_\star \left(B \star v \right)\,,
&&\bar{S}_{\dot \alpha}\star \bar{S}^{\dot \alpha}=2 \bar{f}_\star (B \star \bar{v}\Gamma)\,, 
&&[S_\alpha, \bar{S}_{\dot \alpha}]_\star=0\,,
\label{masterfull4}
\end{align}
\end{subequations}
and the gauge transformations take the form
\beq
\label{gauge}
\delta W=d \epsilon-\left[W, \epsilon \right]_\star\,,\quad
\delta B=\epsilon \star B-B \star \, \pi\left(\epsilon\right)\,,\quad
\delta S_\alpha=\left[\epsilon , S_\alpha \right]_\star \,,\quad
\delta \bar{S}_{\dot \alpha}=\left[\epsilon , \bar{S}_{\dot \alpha} \right]_\star\,.
\eeq

\subsection{Spin-Statistics Theorem}
The construction of the superfields above is based on bosonic variables $Y,Z$ 
carrying half spin and fermionic variables $\vartheta^i$ which are scalars. To 
satisfy the spin-statistics theorem we must project on to half of the components of 
all the fields such that the number of $Y,Z$ is equal to the number of $\vartheta^i$ 
modulo 2. The correct condition to be imposed on the fields is
\begin{align}
W(Y,Z|x,\vartheta)&=\Gamma W(-Y,-Z|x,\vartheta)\Gamma\,,\\
B(Y,Z|x,\vartheta)&=\Gamma B(-Y,-Z|x,\vartheta)\Gamma\,,\\
\epsilon(Y,Z|x,\vartheta)&=\Gamma \epsilon(-Y,-Z|x,\vartheta)\Gamma\,, \label{spinstatistic}
\end{align}
where $\epsilon$ designates a higher-spin gauge parameter. Consistency with the 
equations of motion imposes (this depends on the choice of reality conditions)
\begin{align}
S_\alpha(Y,Z|x,\vartheta)&=-\Gamma S_\alpha(-Y,-Z|x,\vartheta)\Gamma\,,\\
\bar{S}_{\dot \alpha}(Y,Z|x,\vartheta)&=-\Gamma\bar{S}_{\dot \alpha}(-Y,-Z|x,\vartheta)\Gamma\,.
\end{align}
These conditions can alternatively be written using the properties of the kleinians as follows
\beq
\left[v \bar{v} \Gamma, W \right]_\star 
=\left[v \bar{v} \Gamma, B \right]_\star 
=\left[v \bar{v} \Gamma, \epsilon \right]_\star 
=\left\{v \bar{v} \Gamma, S \right\}_\star 
=0\,. \label{superbosonicproj}
\eeq

This ensures that functions which are even functions of the $\vartheta^i$ are even 
functions in $Y$ and $Z$, 
while functions which are odd functions of the $\vartheta^i$ 
are odd functions in $Y$, $Z$. 
This is due to the fact that the matrix $\Gamma$ 
(anti-)commutes with an (odd) even number of $\vartheta^i$. This means that now 
we can consider fermions in the theory while respecting the spin-statistics theorem. 

It should be noted that this supersymmetric extension is different from the ones 
introduced by other authors. In~\cite{n=1234} extensions are constructed out of 
the minimal bosonic theory. Grassmann odd variables are also introduced, but 
some of them have a very distinct role, as they introduce back the fields of odd 
spin. In~\cite{extended}, the fermions are also truncated away but fields of odd 
spin are kept. The fermions are reintroduced with a grading inside the tensored 
matrix algebra. Here the supersymmetric extension is constructed from the 
bosonic theory with fields of all spin, even and odd while keeping the fermions from the start.

\subsection{Generalized Reality Conditions}
\label{nonminimal}
For the equations \eqref{masterfull21}-\eqref{masterfull22} to consistently describe the 
interactions of real massless higher spin fields, one has to impose adequate reality 
conditions on the master fields.%
\footnote{Note that different conventions can be used for the reality conditions. 
This can be consistently tracked down to a different choice for the canonical 
commutation relations between the spinor variables
(sometimes a factor $i$ is 
introduced as in~\cite{99review}). See Appendix~\ref{sec:conventions}.\label{S-convention}}

It should be noted that different signatures imply different isometry groups and by extension 
different higher-spin algebras. Notions of spinors will vary in subtle ways, and so will the 
representation in terms of oscillators. In this work, we are mainly interested in the Lorentzian 
signature, with the convention where we use mostly minus signs. As explained in 
detail in~\cite{realforms}, choosing a different signature puts constraints on the 
consistent reality projections one can choose, and even on the Vasiliev equations, 
especially in the interacting part.
We will now present the reality conditions that we will be using and which are 
identical to the ones in~\cite{triality}.

First we define a natural generalisation of the complex conjugation to the master fields with
\begin{equation}
(y_\alpha)^\dagger=\bar y_{\dot \alpha}\,,
\qquad
(z_\alpha)^\dagger=\bar z_{\dot \alpha}\,,
\qquad
(dz_\alpha)^\dagger=d \bar{z}_{\dot \alpha}\,,
\qquad
(\vartheta^i)^\dagger=\vartheta^i
\end{equation}
For any functions $\Phi,\Theta$ of $(Y,Z|x,\vartheta)$, we then obtain from the 
definition \eqref{starproduct}
\beq
\left(\Phi\star\Theta\right)^\dagger=\Theta^\dagger \star \Phi^\dagger.
\eeq
We then define a second operator $\tau$ (denoted $\iota$ in~\cite{triality}) 
which reverses the order of the $\vartheta^i$ and otherwise acts by
\beq
\tau \left[\Phi(Y,Z,dZ|x,\vartheta) \right] 
=\Phi(iY,-iZ,-idZ|x,\tau[\vartheta])\,.
\eeq
In this way, for any two functions $\Phi,\Theta$ of $(Y,Z|x,\vartheta)$, we obtain
\beq
\tau \left(\Phi \star\Theta \right)=\tau(\Theta) \star \tau(\Phi)\,,
\eeq
where it should be noted that the order of the functions is exchanged. Also we have that 
$\tau\left(\Gamma \right)^\dagger=\Gamma^{-1}=\Gamma$.

Following \cite{triality} we impose the {\em non-minimal} reality conditions
\beq
\tau(W)^\dagger=-W\,,
\qquad
\tau(S)^\dagger=-S\,,
\qquad
\tau(B)^\dagger=\bar{v} \star B \star \bar{v} \Gamma=\Gamma v \star B \star v. \label{susygenproj}
\eeq
This differs from the {\em minimal} reality conditions where in addition to the condition above, 
the fields are also assumed to be invariant under complex conjugation alone.

\subsection{Extended Higher-spin Symmetry}
One convenient way to add degrees of freedom to the theory is to tensor the higher-spin 
algebra with another algebra. There are some restrictions as to the nature of 
the algebras one can choose, which were discussed very early on by Konstein and 
Vasiliev in~\cite{extended}. This procedure amounts to promoting all the fields of the 
theory to matrices%
\footnote{Complex conjugation is replaced by the usual hermitian conjugation with 
respect to the matrix indices.} 
$\Phi(Y,Z|x,\vartheta)\rightarrow \Phi^{\ j}_{i}(Y,Z|x,\vartheta)$, where $i,j=1\dots M$. 
One can choose $U(M)$ as the tensored matrix algebra, and this results in 
having the Maxwell field becoming a $U(M)$ gauge field. This extension is 
called in~\cite{triality} Vasiliev's theory with $U(M)$ Chan-Paton factors.

\subsection{Higher-Spin Holography}
\label{sec:holography}

Higher spin theories provide a realization of the holographic principle, 
as first introduced in~\cite{Klebanov} and \cite{Sezgin} for the $4$-dimensional 
case. We present here a brief review and refer the reader to the original 
papers and \cite{holographyreview,holography,triality} for more details.

As is usual in the $AdS_{d+1}$/CFT$_d$ correspondence, fields in the bulk 
are dual to operators in the boundary theory. In particular for a scalar field 
of mass $m$, the dual operator has dimension
\beq
\label{m-Delta}
\Delta_{\pm}=\frac{d\pm \sqrt{d^2+4m^2 \lambda^2}}{2}\,.
\eeq
This can be seen from the solution of the Klein-Gordon equation, 
which near the boundary of $AdS$ ($r\to0$ in the metric \eqn{Adsmetric}) 
takes the form
\beq
C(r,x)={a \over r^{\Delta_-}}+{b \over r^{\Delta_+}}+\dots 
\eeq
For most fields the dual operator has dimension $\Delta_-$, but for small enough 
$m$, there is an ambiguity, where 
enforcing $b=0$ corresponds to a dual theory an operator of dimension $\Delta_-$ 
while the $a=0$ boundary conditions gives a theory with operator of 
dimension $\Delta_+$.

Furthermore, generic choices of boundary conditions will break the higher spin symmetry. 
It is argued in~\cite{triality} that for fields with spins higher than 
$3/2$, generic boundary conditions preserve the higher-spin symmetry.\footnote{There is however
an unresolved puzzle concerning this statement, for more details see  the end of page 34 
of \cite{triality}. We thank S. Giombi for bringing this point to our attention.}
The problem lies with the fields of spin $0$, $1/2$ and $1$.

In the minimal parity preserving bosonic theory there is one real 
massless field for every even spin and in addition 
one real scalar field with mass $m^2=-2\lambda^{-2}$. Plugging this mass 
and $d=3$ in \eqn{m-Delta}, we find that the dual operator has $\Delta_-=1$ and $\Delta_+=2$. 
This leads to the four simplest holographic duals with either choice of dimension and with the 
phase $\theta(X)$ in the interaction term $f(X)=1+Xe^{i\theta(X)}$ equal to a constant $0$ or $\pi/2$. 
These four choices are dual to either the free or critical, bosonic or fermionic $O(N)$ vector models.

This can be generalized to the case of non-minimal theories, which 
contain fields of all non-negative integer spins. Then one has to take care of the 
fact that the theory contains a gauge field of spin $1$, whose boundary conditions 
will be severely constrained if higher-spin symmetry is to be preserved.
In particular, the case we shall focus on in this paper, of supersymmetric parity 
violating theories was studied in~\cite{triality}. It was argued that the introduction of a 
parity breaking phase $\theta(X)=\theta_0 $ corresponds to gauging the vector model and 
including a Chern-Simons term at level $k$ such that $\theta_0=N/k$ is the 
't~Hooft coupling.  We will return to the question of the boundary 
conditions on the scalar and vector fields in Section~\ref{sec:boundary}.

\section{Embedding Bosonic Solutions}
\label{sec:embedding}

The equations of motion of the supersymmetric higher spin theory involve all the 
the fields of the theory including the fermionic fields and several copies of the bosonic ones. 
A simple organizing principle is to separate the fields into eigenstates of the chirality operator 
$\Gamma=i^{\frac{n(n-1)}{2}} \vartheta^1 \vartheta^2 \dots \vartheta^n$
\bal
\label{decomposition}
W&=\Gamma^+W_++\Gamma^-W_-\,,\\
B&=\Gamma^+B_++i\Gamma^-B_-\,,\qquad
\Gamma^\pm=\frac{1\pm\Gamma}{2}\,,\\
S&=\Gamma^+S_++\Gamma^-S_-\,,\\
\eal
The extra $i$ in the decomposition of $B$ is due to the fact that it is in the twisted adjoint 
representation and it will simplify the equations of motion and reality conditions below. 
Each of the fields $\Phi_\pm$ are superfields which are given by an expansion in $\vartheta^i$ 
restricted of course to half the possible terms. Each of those superfields can then be further 
separated into bosonic and fermionic parts, where the bosonic part has the same 
matter content as the field of the bosonic theory with $U(2^{n/2-1})$ Chan-Paton 
factors (generated by $\Gamma^+\vartheta^{i_1}\cdots\vartheta^{i_{2k}}$). 
The cases discussed below are those with $n=2,4,6$ and hence Chan-Paton 
factors $U(1)$, $U(2)$ and $U(4)$ respectively.

When considering classical solutions we can set all the fermionic fields to zero and 
restrict to the two sets of bosonic fields in $W_\pm$, $B_\pm$ and $S_\pm$. The main 
observation is that the projectors $\Gamma^\pm$ either act trivially or annihilate these 
bosonic fields leading to very simple equations of motion. The flatness equation for $W$ 
\eqn{masterfull1} becomes
\beq
\Gamma^\pm(dW-W\wedge_\star W)
=\Gamma^\pm W_\pm
-\Gamma^\pm W_\pm\wedge_\star\Gamma^\pm W_\pm
=\Gamma^\pm\left(dW_\pm-W_\pm\wedge_\star W_\pm\right).
\eeq
The equations project onto separate bosonic equations for $W_+$ and $W_-$. 
This holds also for the equations (\ref{masterfull2})-(\ref{masterfull5}).

Equations \eqn{masterfull4} require a bit more care, due to the explicit appearance of 
$\Gamma$ in them. They become
\bal
S_{+\alpha} \star S_+^\alpha&=2 f_\star \left(B_+\star v \right)\,,
\quad
&\bar{S}_{+\dot \alpha}\star \bar{S}_+^{\dot \alpha}&=2 \bar{f}_\star (B_+\star \bar{v})\,, 
\quad
&[S_{+\alpha}, \bar{S}_{+\dot \alpha}]_\star&=0\,,\\
S_{-\alpha} \star S_-^\alpha&=2 f_\star \left(iB_-\star v \right)\,,
\quad
&\bar{S}_{-\dot \alpha}\star \bar{S}_-^{\dot \alpha}&=2 \bar{f}_\star (-i B_-\star \bar{v})\,, 
\quad
&[S_{-\alpha}, \bar{S}_{-\dot \alpha}]_\star&=0\,.
\eal
The equation for $S_+, \bar S_+$ are the usual equations as in the bosonic theory, but 
the equation for $\bar S_-$ is different due to the extra sign in $\bar f$. 
Recall that $f(X)=1+Xe^{i\theta(X)}$. If we assume%
\footnote{This is not a very strong assumption, since it is known that $\theta(X)$ can be 
written as a power series in $X^2$ but mostly assumed to be $X$ independent, 
for which a new argument was proposed in~\cite{vasiliev-talk}.} 
that $\theta$ is a power series in $X^4$, then
\beq
f_*(iB_-)=1+B_-\star e_\star^{i(\theta(B_-)+\pi/2)}\,,
\qquad
\bar f_*(-iB_-)=1+B_-\star e_\star^{-i(\theta(B_-)+\pi/2)}\,,
\eeq
Thus the equation for $S_-,\bar S_-$ is the same as the bosonic equation with 
a shift in the phase $\theta(X)\to\theta(X)+\pi/2$.

As mentioned before, 
the simplest two choices for $f$ are given by $\theta(X)=0$ and $\theta(X)=\pi/2$. These two 
examples are the only ones which do not break parity. They correspond respectively 
to the so-called type-$A$ (with parity-even $B$-field) and type-$B$ 
(with parity-odd $B$-field) theories. Classical solutions of a parity invariant supersymmetric 
theory can therefore incorporate solutions of both the type-$A$ and type-$B$ bosonic 
theories.

\subsection{Matrix Factors} 
As shown, any solution of the supersymmetric equation can be written in terms of 
solutions of the bosonic theory with $U(M)$ Chan-Paton factors. We would like to discuss here 
how to construct a solution to the bosonic theory with Chan-Paton factors, based on the 
solutions to the abelian theory. Unlike the previous discussion, where we found all 
possible solutions, in this case we will take a simple ansatz and will not find the most 
general solution. The ansatz is
\bal
\label{matrixfactor}
W_i{}^j(Y,Z|x)&=w_i{}^j \tilde W(Y,Z|x)\,,\\
B_i{}^j(Y,Z|x)&=b_i{}^j \tilde B(Y,Z|x)\,,\\
S_{\alpha,i}{}^{j}(Y,Z|x)&=s_i{}^j \tilde S_\alpha(Y,Z|x)\,,\\
\bar S_{\dot\alpha,i}{}^j(Y,Z|x)&=\bar s_i{}^j \tilde{\bar{S}}_{\dot\alpha}(Y,Z|x)\,.\\
\eal
We take $w$, $b$, $s$ and $\bar s$ to be matrix pre-factors for the classical solution 
given by $\tilde W(Y,Z|x)$, $\tilde B(Y,Z|x)$, $\tilde S_\alpha(Y,Z|x)$ and $\tilde{\bar{S}}_{\dot\alpha}(Y,Z|x)$. 
We deduce the reality conditions that apply on the matrix factors
\beq
\label{matrixreal}
w^\dagger=w \, , \, b^\dagger=b \, , \, s^\dagger=\bar s.
\eeq
The equations of motion \eqn{masterfull1}-\eqn{masterfull4} will be solved for this very 
general ansatz only if all the terms in each equation are proportional to each other%
\footnote{We assume here that the interacting phase is a constant.}
\bal
w^2&=w \,,\qquad
&w b&=b w=b\,,\qquad
&ws&=sw=s \,,\qquad
&w\bar s&=\bar sw=\bar s \,,\\
s^2&=\bar s^2=b\,,\qquad
&s\bar s&=\bar ss \,.
\eal
All the matrices commute, so can be simultaneously diagonalized. 
$w$ is a projector which acts trivially on an $m$ dimensional subspace and 
annihilates an $M-m$ dimensional one. $s$ has 
to vanish on the same space as $w$ and may have up to $m$ non-zero eigenvalues. 
$b$ is then determined by $b=s^2$. 

When the entries in the matrices are all made of a unique solution we saw that 
the matrices can all be simultaneously diagonalized. In that case there is no 
reason to restrict them to be proportional to a single solution of the bosonic theory 
and we can construct a more general solution made of different solutions along the diagonal.

\subsection{Supersymmetry Invariance}
Global symmetries 
(including supersymmetry) of a classical solution of higher spin theory
are represented by 
gauge symmetries which leave the
solution invariant. The prime example is 
given by the global symmetries of the vacuum solution \eqref{vac1}-\eqref{vac2}. 
First, the equation $\delta S_0=0$ tells us that the gauge parameter $\epsilon$ is 
independent of the $Z$ variables. The $B$ field transforms homogeneously and 
thus gives no additional constraints. Finally with the equation $\delta W_0=0$ we 
conclude that the global symmetries of the vacuum are generated by the gauge 
parameter $\epsilon(Y|x, \vartheta)$ satisfying
\beq
d \epsilon-[W_0, \epsilon]_\star=0. \label{HSsymmetries}
\eeq
There may be solutions to this equation 
of the form 
$\epsilon(Y|x,\vartheta)=R_{ij} \vartheta^i \vartheta^j$ for $i \neq j$. From 
\eqref{HSsymmetries} and the reality condition 
$\tau(\epsilon(\vartheta))^\dagger=-\epsilon(\vartheta)$ we deduce that 
$R_{ij}$ is a real space-time independent parameter. These solutions 
correspond to the generators of the R-symmetry of the theory, which will 
be broken in later sections upon introduction of boundary conditions.

The second type of solutions corresponds to gauge parameters that are 
proportional to $Y$. 
These are spinorial, and hence supersymmetry generators. Let us write these gauge parameters 
as
\beq
\epsilon(Y|x,\vartheta)=\Xi_{\alpha}(x,\vartheta)y^\alpha
+i \bar{\Xi}_{\dot \alpha}(x,\vartheta) \bar{y}^{\dot \alpha} \,, 
\label{n4basis}
\eeq
where 
the reality condition 
sets 
$\bar\Xi_{\dot\alpha}=\Xi_\alpha^\dagger$ 
and they are odd functions of the $\vartheta^i$ due to the truncation \eqref{spinstatistic}. 
If we replace the components of $W_0$ by the vierbein and connection $1$-form of 
the $AdS_4$ metric using \eqn{vac1}-\eqref{vac2}, we can rewrite \eqref{HSsymmetries} as
\beq
\tilde \nabla \begin{pmatrix}\Xi\\\bar\Xi\end{pmatrix} 
\equiv \left(d-\frac{i}{2} \omega_{ab} \gamma^{ab}+\frac{i}{\sqrt 2} h_a\gamma^a \right)
\begin{pmatrix}\Xi\\\bar\Xi\end{pmatrix} 
=0 \,,
\label{AdSkillingspinor}
\eeq
where we split the equation into its $y^\alpha$ and $\bar y^{\dot\alpha}$ components. 
This is nothing but the Killing spinor equation in $AdS$ background.

For $AdS_4$, the solution to the spinor part of \eqref{AdSkillingspinor} is well known \cite{freedman}, 
and is parametrized by four independent constants, as 
we summarize 
in Appendix~\ref{sec:KillingSpinorGlobal}. 
Since there are $2^{n-1}$ odd functions 
of $\vartheta^i$, there are a total of $2^{n+1}$ real supersymmetry parameters 
for the $AdS_4$ background.

We can thus express the parameter $\epsilon$ \eqref{n4basis} as linear combinations of the Killing spinors of $AdS_4$
$\psi^I=\psi^I_\alpha y^\alpha + i\bar \chi^I_{\dot\alpha}\bar y^{\dot\alpha}$
\beq
\label{lincomb}
\epsilon(Y|x,\vartheta) = \psi^I(Y|x) \xi^I(\vartheta)\, , 
\eeq
with $I=1,2, \bar 1,\bar 2$.
Due to the reality condition, the various $\xi^I$ are related through
\beq
\label{daggerprop}
\big( \xi^i\big)^\dagger = \xi^{\bar i}\,,
\eeq
with $i=1,2$.
To see this, we write
\beq
\tau\left( \psi^i \right)^\dagger
= \tau\left( \psi^i_\alpha y^\alpha + i\bar \chi^i_{\dot\alpha}\bar y^{\dot\alpha} \right)^\dagger
=- i \left( \bar \psi^i_{\dot \alpha} \bar y^{\dot \alpha} - i \chi^i_{\alpha} y^{\alpha} \right) 
= - \left( \chi^i_{\alpha} y^{\alpha} + i \bar \psi^i_{\dot \alpha} \bar y^{\dot \alpha} \right)
= - \psi^{\bar i} \, , 
\eeq
where we have used the properties \eqref{conjugationprop}, so that
\beq
\tau\big(\xi^i \psi^i + \xi^{\bar i} \psi^{\bar i} \big)^\dagger 
= - \big(\xi^i\big)^\dagger \psi^{\bar i} - \big(\xi^{\bar i} \big)^\dagger \psi^{i}
= -\big(\xi^i \psi^i + \xi^{\bar i} \psi^{\bar i} \big)
\eeq
where the last equality comes from the reality condition.

In 
the remainder of 
this paper we will 
study the Didenko-Vasiliev solution, which as we show in Appendix~\ref{sec:DV}, 
has $W=W_0$ --- the same connection as the vacuum $AdS$ solution. These solutions have 
non-trivial values for the $B$ and $S$ fields, so to check supersymmetry we will therefore 
have to impose that they are invariant under the gauge transformations \eqn{n4basis}. 
Plugging this into \eqn{gauge} gives
\begin{align}
\psi^I\star B\,\xi^I b - B \star\pi(\psi^I)\, b\xi^I &=0 \label{B-susy}\,, \\
\psi^I\star S\,\xi^I s -S \star \psi^I \, s\xi^I &=0 \label{S-susy}\,.
\end{align}
The solutions to these equations determine the bulk supersymmetries of the solutions and 
further conditions need to be checked to verify that they are not broken by the boundary 
conditions, see Section~\ref{sec:boundary}.

\section{Supersymmetric Embedding of the Didenko-Vasiliev Solution}
\label{sec:DV-embed}

We turn our attention now to the Didenko-Vasiliev (DV) solution of the bosonic 
theory \cite{DV}, whose construction is outlined in Appendix~\ref{sec:DV}. 
The presentation in the appendix goes beyond that in \cite{DV}, first we 
generalize it to a bosonic theory with arbitrary parity 
breaking phase $\theta$. 
Second, we obtain a striking simplification by 
choosing a particular Killing vector and a particular Lorentz frame such that the quantity $\kappa_{\alpha \beta}/r$ defined in \eqref{KillingDV} is space-time independent.
We find that the $W$ master 
field simply takes the $AdS_4$ background value $W_0$. The full 
solution is given by \eqref{sol1}-\eqref{sol4}.

In the original paper 
\cite{DV} this solution was also embedded into a supersymmetric theory 
with the same amount of supersymmetry as the one with $n=2$ (though 
in a 
different formalism) and it was 
checked that it preserves $1/4$ of the supercharges in the bulk.

We 
implement now  
our formalism from Section~\ref{sec:embedding} and 
find all such embeddings of the bosonic solution into the supersymmetric theory with 
$n=2,4$. We then check under what conditions those solutions are BPS and 
whether they are compatible with boundary conditions conjectured in~\cite{triality} 
to be dual to different supersymmetric $3$d-theories.

We saw in Section~\ref{sec:embedding}, that one can diagonally embed any bosonic solution 
into the supersymmetric theory with the restriction that the solution in the block projected to 
by $\Gamma^-$ is in the theory with the phase shifted by $\pi/2$. As mentioned above, the 
DV solution can be easily adapted to arbitrary phase. To find BPS embeddings 
we shall consider two solutions with different Killing matrices differing by sign; $K$ and $-K$ 
\cite{DV}.

In a basis where all the matrices are diagonal our embedding 
has 
$W=W_0$, or 
$w=1$ in \eqn{matrixfactor} (the case where $w$ has vanishing eigenvalues does not 
seem particularly interesting). The fields $B$ and $S$ 
are 
diagonal with every entry 
involving either of the Killing matrices $\pm K$ and a continuous 
parameter $b$.

To be specific, the ansatz is (see Appendix~\ref{sec:DV} for the ingredients of the solution)
\bal
\label{SUSYDVansatz}
W(Y,Z|x)&=W_0 \,,\\
B(Y,Z|x)&= b \left[ \eta_p F_{K} +\eta_m F_{-K}\right] \star \delta (y) \,,\\
S_{\alpha}(Y,Z|x)&=z_\alpha + s \big[ \eta_p F_{ K} \sigma_\alpha(a, K|x) 
+ \eta_m F_{- K} \sigma_\alpha(a, -K|x) \big] \,,\\
\bar S_{\dot\alpha}(Y,Z|x)
&=\bar{z}_{\dot \alpha}+ \bar s \big[ \eta_p F_{ K} \bar \sigma_{\dot \alpha}(\bar a, K|x) 
+\eta_m F_{-K} \bar \sigma_{\dot \alpha}(\bar a, -K|x) \big]\,,
\eal
where $\eta_p,\eta_m$ are orthogonal projectors. The matrices (including $\eta_{p}$, $\eta_m$) 
are diagonal, and we also define
\bal
\label{bssbar}
b &\equiv  b_+ +   i b_- \equiv \diag (b_{+,1},b_{+,2}, i b_{-,1}, i b_{-,2}) \, , \\
s &\equiv e^{i \theta_0}s_{+} + e^{i (\theta_0 +\pi/2)} s_{-} \, , \\
\bar s &\equiv e^{-i \theta_0}\bar s_{+} + e^{-i (\theta_0 +\pi/2)} \bar s_{-} \, , 
\eal
where the $\pm$ subscript indicates as usual the fact that the matrix is an eigenstate of 
$\Gamma^\pm$, and where all the parameters in $b$ are real.
The 
equations of motion then impose the following constraints on the matrix factors
\beq
s_{\pm}=b_{\pm} \,,\qquad
bs=sb \,,\qquad
s\bar s=\bar s s \,,
\eeq
where we have omitted the equations obtained from the barred sector which can be deduced 
from the unbarred sector by complex conjugation and using the reality conditions which read 
for the matrix factors in \eqref{SUSYDVansatz}
\beq
\label{realityprop}
b_{\pm}^\dagger = b_{\pm} \, , 
\qquad 
s^\dagger = \bar s.
\eeq
One can of course rotate this solution to a different basis, where it is not diagonal.

\subsection{The BPS Equations for the DV Solution}
Following the discussion in 
Section~\ref{sec:embedding}, our ansatz \eqref{SUSYDVansatz} 
is BPS if it solves the equations 
\eqref{B-susy} and \eqref{S-susy} for 
non-trivial gauge parameters of the form
\beq
\epsilon(Y|x,\vartheta) = \psi^I(Y|x) \xi^I(\vartheta)\, , 
\eeq
where $\psi^I$ are the Killing spinors of $AdS_4$ given in Appendix~\ref{sec:KillingSpinorGlobal}.

Note that the DV solution is based on a black hole solution of 
supergravity with a Killing matrix $K_{AB}$ discussed in Appendix~\ref{sec:symmetries}. 
The $B$ and $S$ fields are proportional to the star product projectors $F_K$ \eqn{FK}. 
In particular $F_K$ projects the four $Y_A$ coordinates to a two-dimensional subspace
\beq
(\Pi_{-A}^{\ \ \ B}Y_B) \star F_K=F_K \star (\Pi_{+A}^{\ \ \ B}Y_B)=0\,,
\qquad
\Pi_{\pm AB}=\half \left(\epsilon_{AB} \pm i K_{AB} \right)
\eeq
In fact, since $K_{AB}$ is a bilinear in the $AdS_4$ Killing spinors, 
$F_K$  
projects the four Killing spinors onto 
a 
two-dimensional subspace. We discuss in detail the properties of the projector 
$\Pi_\pm$ and the Killing matrix $K_{AB}$ \eqref{KillingDV} and how it is related to the 
Killing spinors in 
Appendix~\ref{sec:DV-Killing}. Here we write for convenience the 
following properties we will be using. We have
\beq
\psi^{i}=\Pi_{+}\psi^{i}\,, 
\qquad 
\psi^{\bar i}=\Pi_{-}\psi^{\bar i}
\, ,
\eeq
which gives
\beq
\label{proprep}
\psi^{i}\star F_K=F_K\star\psi^{\bar i}=0\,,
\qquad
F_K\star\psi^{i}=2 F_K\psi^{i}\,, 
\qquad 
\psi^{\bar i}\star F_K=2 \psi^{\bar i}F_K\, .
\eeq
for $i=1,2$.
Using these properties, we can simplify the BPS equation \eqref{B-susy} as follows
\beq
\psi^iF_{-K} \xi^i \eta_m b
+\psi^{\bar i} F_K \xi^{\bar i}\eta_p b
-F_K \psi^i \eta_p b\xi^i
-F_{-K} \psi^{\bar i} \eta_m b \xi^{\bar i} =0\,.
\eeq
This equation is satisfied if 
\beq
\label{simplifiedBPSeq}
\xi^i \eta_m b = \eta_p b\xi^i =0\,.
\eeq
Such an equation is easy to solve, as it is purely algebraic. 
We take $b$ to be generic, with no zero eigenvalues on the space that $\eta_p+\eta_m$ projects 
onto (the value of $b$ on the orthogonal space is not important, as it is always projected out). 
For each nonzero entry in $\eta_m$, the corresponding column of $\xi^i$ has to vanish and 
for each nonzero entry in $\eta_p$, the corresponding line of $\xi^i$ has to vanish. 
Clearly, both the $i$-th column and the $i$-th line for a fixed $i$ cannot be crossed out 
simultaneously, since $\eta_p \eta_m=0$.
The BPS equation \eqref{S-susy} is then automatically 
satisfied, as properties similar to \eqref{proprep} are satisfied when $F_{\pm K}$ is replaced 
by $F_{\pm K} f(a)$, where $f(a)$ is any holomorphic (or anti-holomorphic) function of $a$ 
(see Appendix~\ref{sec:DV} and \eqref{generalprojphi}).

\subsection{$n=2$}

The supersymmetric theory with $n=2$ is particularly simple in the sense that the 
two $\Gamma^\pm$ subspaces are of rank 1. We thus have two independent eigenvalues. If both are 
non-zero, we have $\eta_p + \eta_m =1$. Then we see that the only 
configuration with a non-trivial $\xi^i$ which still satisfy \eqref{simplifiedBPSeq} 
is (up to exchanging the role of $F_K$ and $F_{-K}$)
\beq
\label{n2halfbpsb}
\eta_p=\diag(1,0)\,, \quad \eta_m = \diag(0,1) 
\eeq
and preserves the supersymmetry with $\vartheta^i$-content
\beq
\label{n2halfbpsxi}
\xi^i = \begin{pmatrix} 0 & 0 \\ * & 0 \end{pmatrix} \,,
\eeq
and this corresponds to a half-BPS configuration, where in this section we 
include all the supersymmetries 
of the bulk for this counting. This will change when we 
impose boundary conditions 
which break some supersymmetries 
in the next section. 
In particular, if both eigenvalues are to be non-zero, this means that the 
$B$-field cannot contain only $F_K$ (or only $F_{-K}$) and still be BPS.
If we impose only one of the two eigenvalues to be non-vanishing, then 
we only need to cross either a line or a column, and the preserved supersymmetry 
will have a $\vartheta^i$-content where only one 
off-diagonal entry is non-zero, leading again to half-BPS configurations.

Finally, the case where both eigenvalues of $b$ vanish simply corresponds 
to the $AdS_4$ vacuum which is maximally supersymmetric.

\subsection{$n=4$}
We now turn to the $n=4$ case, where $b$ has four eigenvalues. Again, starting 
with the configuration where none of the eigenvalues vanish, the most 
supersymmetric configuration is given by (again, up to the exchange of the 
role of $F_K$ and $F_{-K}$)
\beq
\label{n41/2BPS}
\eta_p = \diag (1,1,0,0)\,,
\qquad
\eta_m=\diag(0,0,1,1)\,,
\eeq
as it preserves the supserymmetries with $\vartheta^i$-content
\beq
\label{n41/2BPSxi}
\xi^i = \begin{pmatrix} 0 & 0 & 0 & 0 \\0 & 0 & 0 & 0 \\ * & * & 0 & 0 \\ * & * & 0 & 0 \end{pmatrix},
\eeq
which gives a half-BPS configuration. There are other configurations where $B$ has four 
non-vanishing eigenvalues, which preserve 
$1/4$ of the supersymmetries. In a basis where $b$ is diagonal they are
\beq
\label{n41/4BPS}
\eta_p = \diag ( 1,0,1,0 )\,,
\qquad
\eta_m=\diag(0,1,0,1)\,,
\qquad \xi^i 
= \begin{pmatrix} 0 & 0 & 0 & 0 \\0 & 0 & * & 0 \\ 0 & 0 & 0 & 0 \\ * & 0 & 0 & 0 \end{pmatrix}.
\eeq

Configurations 
where $\eta_p+\eta_m$ is not the identity 
will generically preserve 
more supersymmetries. If we consider 
only three non-vanishing 
eigenvalues, we can obtain configurations 
that preserve a different amount of supersymmetry, such as the $3/8$-BPS
\beq
\label{n43/8BPS}
\eta_p = \diag (1,0,0,0 )\,,
\qquad
\eta_m=\diag(0,1,0,1)\,,
\qquad 
\xi^i = \begin{pmatrix} 0 & 0 & 0 & 0 \\0 & 0 & * & 0 \\ * & 0 & 0 & 0 \\ * & 0 & 0 & 0 \end{pmatrix}.
\eeq

Now turning on to 
only two non-vanishing eigenvalues, 
we obtain the following 
two cases. The $1/2$-BPS
\beq
\label{n41/2'BPS}
\eta_p = \diag ( 1,0,0,0)\,,
\qquad
\eta_m=\diag(0,1,0,0)\,,
\qquad
\xi^i = \begin{pmatrix} 0 & 0 & 0 & 0 \\0 & 0 & * &* \\ * & 0 & 0 & 0 \\ * & 0 & 0 & 0 \end{pmatrix},
\eeq
and the $5/8$-BPS
\beq
\label{n45/8BPS}
\eta_p = \diag (1,0,0,0 )\,,
\qquad
\eta_m=\diag(0,0,0,1)\,,
\qquad
\xi^i = \begin{pmatrix} 0 & 0 & 0 & 0 \\0 & 0 & * & 0 \\ * & * & 0 & 0 \\ * & * & 0 & 0 \end{pmatrix}.
\eeq

Finally if the $B$-field has only one non-vanishing eigenvalue, we preserve 
$3/4$ of the supercharges
\beq
\label{n43/4BPS}
\eta_p = \diag ( 1, 0,0,0)\,,
\qquad
\eta_m=\diag(0,0,0,0)\,,
\qquad
\xi^i= \begin{pmatrix} 0 & 0 & 0 & 0 \\0 & 0 & * & * \\ * & * & 0 & 0 \\ * & * & 0 & 0 \end{pmatrix}.
\eeq

\section{Compatibility of Solutions with Boundary Conditions}
\label{sec:boundary}

Thus far we have studied the embeddings of the DV solution into different supersymmetric 
extensions of higher spin theory with $n=2,4$. We now turn to study the 
asymptotics of the solutions. It was proven in~\cite{maldacena} that theories with higher spin 
symmetry in 4d are holographically dual to free field theories. To describe interacting theories 
one needs to break the higher spin symmetry, which can be achieved by imposing different 
boundary conditions on the fields.

The boundary conditions for the supersymmetric theories were studied in~\cite{triality} 
where the fields of spin $(0,\half,1)$ can break higher-spin symmetries as well as 
supersymmetry. Furthermore, different boundary conditions were related there to different 
supersymmetric $3$d CFTs. To study the classical solutions we need to concern ourselves only 
with the boundary conditions of the bosonic fields --- the scalar and vector components of $B$, 
whose fall-off near the boundary of $AdS$ can take the form
\begin{align}
B^{(0)}&=\frac{1}{r} \left(\Gamma^+\cos \gamma+i\Gamma^-\sin \gamma \right) \tilde{f}_1
+\frac{1}{r^2} \left(\Gamma^-\cos \gamma+i\Gamma^+\sin \gamma \right) \tilde{f}_2 
+O\Big(\frac{1}{r^3}\Big),
\label{scalarbound}\\
B^{(1)}&=\frac{1}{r^2} \left[e^{i \beta} F_{\alpha\beta} y^\alpha y^\beta
+\Gamma e^{-i\beta} \bar{F}_{\dot\alpha\dot\beta}\bar{y}^{\dot\alpha}\bar{y}^{\dot\beta}\right] 
+O\Big(\frac{1}{r^3}\Big),
\label{onebound}
\end{align}
where $\tilde{f}_{1,2}$, $F$, $\bar{F}$ are functions%
\footnote{$F$ and $\bar F$ are of course related via the reality condition imposed on $B$.} 
of space-time and of the $\vartheta^i$ and the choice of asymptotics was imposed in 
\cite{triality} through them.

All our solutions are based on the DV solution and the asymptotics will be governed by the 
expansion of \eqref{SUSYDVansatz} using%
\footnote{One should note that the DV solutions are obtained in a gauge in which they take the
deceivingly simple form of a perturbation around the $AdS_4$ vacuum. However this gauge is different
from the so-called \emph{physical} gauge used in \cite{triality} 
for which there is a prescription to linearise the theory
and to make contact with Fronsdal's theory. The complete form of the DV solutions are not known in
this physical gauge. At first order in $b$ however, the master fields $B$ coincide in these two gauges,
and this allows us to use \eqref{scalarbound} with \eqref{Bexpansion} for our discussion. Starting from the next order in $b$, due to the non-linearity of the equations of motion there will most likely be additional terms both in the $1/r$ and $1/r^2$ branches.}
\eqn{sol3}
\beq
\label{Bexpansion}
B=\frac{4}{r}\,b(\eta_p + \eta_m)
-\frac{2i}{r^2} \, b(\eta_p - \eta_m)\left({\kappa_{\alpha \beta} \over r} y^\alpha y^\beta 
+ {\kappa_{\dot \alpha \dot \beta} \over r} \bar{y}^{\dot \alpha} \bar{y}^{\dot \beta} \right)
+O(Y^4)\, .
\eeq
In particular we see that the scalar piece does not have a $1/r^2$ 
component, so we'll have to set $\tilde f_2=0$ throughout, and will rename $\tilde f_1$ 
as $\tilde f$.
 
In this section we 
study the subsets of supersymmetries preserved by the different 
boundary conditions proposed in~\cite{triality} and see whether they can be preserved 
by the field configurations satisfying the equations of motion. In this sense, the counting 
of the preserved supersymmetries 
differs 
from the one of the previous section. 
The boundary 
conditions correspond to specific choices for $(\beta, \gamma, \tilde f_{1,2})$ which 
in turn constrain our ansatz \eqref{SUSYDVansatz}. One important point to note 
here is that the following equation
\beq
\label{btildef}
4 b(\eta_p + \eta_m) = \left(\Gamma^+ \cos\gamma + i \Gamma^-\sin\gamma \right) \tilde f
\eeq
obtained from \eqref{scalarbound} and \eqref{Bexpansion} tells us that the boundary 
conditions that are expressed through constraints on $\tilde f$ 
generically impose some relation between the two chiral parts $b_\pm$ of $b$.

The constraints imposed on $(\beta, \gamma, \tilde f_{1,2})$ are expressed in 
\cite{triality} in terms of commutation relations with the $\vartheta^i$. To make 
the link with the previous section explicit, we choose a specific matrix representation 
for the $\vartheta^i$, that we write down in Appendix~\ref{sec:Matrices}. 

In the previous section we worked in a basis where the master fields were diagonal. 
In this section we will choose a particular basis for the supersymmetry generators 
that preserve the boundary conditions (and $\beta$, $\gamma$, $\tilde f$), so 
we 
need 
in principle 
to repeat the analysis of Section~\ref{sec:DV-embed} in the specific bases. 
That requires to 
solve the BPS equation \eqref{simplifiedBPSeq} for 
a general linear combination of the supersymmetries preserved by 
the imposed boundary conditions.
In practice, we were able to choose the supersymmetry generators in the examples 
below such that they are compatible with the basis of solutions in Section~\ref{sec:DV-embed}.

\subsection{$\mathcal{N}=2$}
 
We 
start with the bulk theories which are conjectured to be dual to 
$\mathcal N=2$ Chern-Simons vector models with matter. We consider two types of boundary 
conditions that are dual to two different such theories.
 
\subsubsection{$SU(2)$ Flavour Symmetry}
The first theory preserves the supersymmetries \eqref{lincomb} with the 
$\vartheta^i$-dependence restricted to $(\vartheta^1,\Gamma \vartheta^1)$.%
\footnote{Here we should note that in \cite{triality}, the supersymmetry generators 
are parametrized by two constant spinors $\Lambda_0$ and $\Lambda_-$ which 
are functions of the $\vartheta^i$. 
Only the structure of $\Lambda_0$
has been explicitly given, and we here assume that $\Lambda_-$ is generated by 
the same $\vartheta^i$ as $\Lambda_0$.}
It is the $n=4$ supersymmetric extended Vasiliev theory with the boundary conditions
\beq
\beta=\gamma=\theta_ 0\,,
\qquad 
\big[\vartheta^1 , \tilde f \big]=0 \,,
\label{boundcon1d}
\eeq
which is conjectured in~\cite{triality} 
to be dual to the $\mathcal{N}=2$ Chern-Simons vector model with two fundamental 
chiral multiplets. The $SU(2)$ flavour symmetry rotating the two chiral multiplets is 
identified with the $SO(3)$ symmetry of rotations in 
$\vartheta^j$, for $j=2,3,4$.

To solve equation \eqn{simplifiedBPSeq} we note that $\xi^i$ is a linear combination of 
$\Gamma^\pm\vartheta^1$. For either of the signs we get the equations
\beq
\Gamma^\mp \eta_m b =\Gamma^\pm \eta_p b =0\,. 
\eeq
Since $\Gamma^\pm$ are orthogonal projectors, non-trivial solutions exist only when 
$\xi^i$ equals either of the two, and not a general linear combination. For example, 
if $\xi^i=\Gamma^+\vartheta^1$, 
then $\eta_m$ is an eigenstate of 
$\Gamma^-$ while $\eta_p$ is an eigenstate of $\Gamma^+$. 
Another way to say this is to note that if we write $\xi^i$ in 
\eqn{n41/2BPSxi} in terms of $\vartheta^i$ (see Appendix~\ref{sec:Matrices}), 
then it is a linear combination of $\Gamma^-\vartheta^i$, 
so in particular preserves $\Gamma^-\vartheta^1$, but none of the other examples will 
also preserve $\Gamma^+\vartheta^1$ (or a different pair $\Gamma^\pm\vartheta^i$, since we 
may be in a different basis). So indeed for the choice of $\Gamma^-\vartheta^1$ 
we find $\eta_p$ and $\eta_m$ as in \eqn{n41/2BPS} which is half BPS.

Next, the second equation in \eqref{boundcon1d} tells us that $\tilde f$ is generated by 
$1$, $\vartheta^2 \vartheta^3$, $\vartheta^2 \vartheta^4$ and $\vartheta^3 \vartheta^4$. 
Since $\tilde f$ cannot be an eigenstate of $\Gamma^\pm$, equation \eqn{btildef} implies 
that $b_+$ and $b_-$ are proportional to each other (with proportionality constant $\tan\theta$) 
and therefore the solution is given by four arbitrary parameters, a generic 
(not necessarily diagonal) $b_+$. 
Lastly, $B^{(1)}$ is not constrained further by \eqref{boundcon1d}.

We conclude that we have found a four-parameter family of half-BPS 
configurations to the theory with boundary conditions \eqref{boundcon1d}.

\subsubsection{$U(1) \times U(1)$ Flavour Symmetry}

In the second case with $\cN=2$ SUSY, the boundary conditions preserve the two supersymmetries generated 
by $(\vartheta^1,\vartheta^2)$ and are given by
\beq
\beta=\theta_0\,,
\qquad \gamma= \theta_0 P_{1,\vartheta^3 \vartheta^4}\, , \qquad
\tilde f \in\spn\left\{1, \vartheta^3 \vartheta^4,\vartheta^3 \vartheta^1, \vartheta^3 \vartheta^2 , \vartheta^4 \vartheta^1, \vartheta^4 \vartheta^2 \right\}\, ,
\label{boundcond2}
\eeq
where $P_{1,\vartheta^3 \vartheta^4}$ projects onto the subspace spanned by 
$1,\vartheta^3 \vartheta^4$. It is conjectured in \cite{triality} that the 
dual boundary
theory is the $\mathcal{N}=2$ Chern-Simons vector 
model with one fundamental and one anti-fundamental chiral matter, 
with $U(1) \times U(1)$ 
flavour symmetry corresponding to the components of the bulk vector gauge field proportional 
to $1,\vartheta^3 \vartheta^4$. The last statement in \eqref{boundcond2} 
means that each monomial generating
$\tilde f$ commutes with either $\vartheta^1$ and/or $\vartheta^2$. 
The projector in $\gamma$ in \eqref{boundcond2} allows us to rewrite 
\eqref{btildef} in the following way
\beq
4b(\eta_p + \eta_m) = \left( \Gamma^+ \cos\theta_0 
+ i \Gamma^- \sin \theta_0 \right) \tilde f (1,\vartheta^3 \vartheta^4) 
+ \Gamma^+ \tilde f' (\vartheta^3 \vartheta^1, \vartheta^3 \vartheta^2 , \vartheta^4 \vartheta^1, \vartheta^4 \vartheta^2) 
\eeq
However, 
$\spn\{\vartheta^3\vartheta^1,\vartheta^3\vartheta^2,\vartheta^4\vartheta^1,\vartheta^4\vartheta^2 \}
=\spn\{\Gamma^\pm\vartheta^3\vartheta^1,\Gamma^\pm\vartheta^3\vartheta^2\}$, 
and the $\Gamma^+$ factor in front of $f'$ restricts its dependence to 
$f'(\Gamma^+ \vartheta^3 \vartheta^1, \Gamma^+ \vartheta^3 \vartheta^2)$. 
We conclude that $b$ has at most four real independent parameters.

To make the link with section \ref{sec:DV-embed}, we note that the $\xi^i$ in 
\eqn{n41/2BPSxi} all involve a chiral projection $\Gamma^-\vartheta^i$. These cannot be 
made out of the pair $\vartheta^1, \vartheta^2$ (or any rotations of them). On the other hand, 
the $\xi^i$ in \eqn{n41/4BPS} is equal to $\vartheta^1-i\vartheta^2$ (for appropriate 
values of the stars), 
so this configuration will be half-BPS (it is easy to see that it does not 
preserve the second supersymmetry).

Imposing \eqref{boundcond2} 
restricts through \eqref{btildef} the parameter space of the solutions. We find 
that $b(\eta_p + \eta_p)$ has two real parameters. 
We 
thus 
find a two-parameter family of half-BPS configuration for the theory with boundary 
conditions \eqref{boundcond2}.

\subsubsection{A One-parameter Family of $\mathcal{N}=2$ Theories}
The boundary conditions just discussed above are in fact a special point in a one-parameter family of $\mathcal{N}=2$ theories preserving the same set of supersymmetries. The boundary conditions are similar to \eqref{boundcond2} with $\gamma$ now having an additional term
\bal
\gamma = \theta_0 P_{1,\vartheta^3 \vartheta^4} + \tilde \alpha P_{\vartheta^2 \vartheta^4,\vartheta^1 \vartheta^4}\, ,
\eal
where $\tilde \alpha$ is a real parameter. By taking $\tilde \alpha$, we recover \eqref{boundcond2},while taking $\tilde \alpha = \theta_0$ yields a theory with supersymmetry enhanced to $\mathcal{N}=3$ described in the next section.
The discussion for preservation of supersymmetry is the same as before. The difference is that \eqref{btildef} now reads
\bal
4 b(\eta_p + \eta_m)=& \left( \Gamma^+ \cos\theta_0 
+ i \Gamma^- \sin \theta_0 \right) \tilde f (1,\vartheta^3 \vartheta^4) 
+ \Gamma^+ \tilde f' (\vartheta^3 \vartheta^1, \vartheta^3 \vartheta^2 ) \nn \\
&+ \left( \Gamma^+ \cos\tilde \alpha 
+ i \Gamma^- \sin \tilde \alpha \right) \tilde f'' (\vartheta^2 \vartheta^4,\vartheta^1 \vartheta^4)
\eal
The situation is similar as to the one for $\tilde \alpha=0$, as we find the same number of parameters. 
\subsection{$\mathcal{N}=3$}

The boundary conditions given by
\beq
\beta=\gamma=\theta_0\,,
\qquad
\tilde f\in\spn\{1,\vartheta^1\vartheta^4,\vartheta^2\vartheta^4,\vartheta^3\vartheta^4\}\,,
\label{boundcond123}
\eeq
preserve the supersymmetries generated by $(\vartheta^1,\vartheta^2,\vartheta^3)$, 
and yield a bulk theory conjectured to be dual to the $\mathcal{N}=3$ Chern-Simons 
vector model with a single fundamental hypermultiplet.

As in the last example, there is no chiral supersymmetry preserved by the boundary 
conditions, which excludes solutions of the type \eqn{n41/2BPS}. But we can look 
at solutions which preserve the same supercharge as in the previous case, 
$\xi^i=\vartheta^1-i\vartheta^2$, which corresponds to \eqn{n41/4BPS}. It is easy to 
verify that no other linear combination of $\vartheta^1$, $\vartheta^2$ and 
$\vartheta^3$ are preserved, so these configurations are $1/3$-BPS.

As before, the second equation in \eqref{boundcond123} reduces the parameter 
space to two real parameters. 
This leads to 
a two-parameter family 
of $1/3$-BPS configuration for the theory with boundary conditions \eqref{boundcond123}.

\subsubsection{A One-Parameter Family of $\mathcal{N}=3$ Theories}
The boundary conditions just discussed are in fact a special point in a one 
parameter family of $\mathcal{N}=3$ theories preserving the same set of 
supersymmetries. The boundary conditions are
\beq
\beta=  \theta_0 (1 - P_\Gamma) + \tilde \beta P_\Gamma \, , \quad 
\gamma = \theta_0 P_1 
+ \tilde \beta P_{\vartheta^1 \vartheta^4,\vartheta^2 \vartheta^4,\vartheta^3 \vartheta^4}\, ,
\quad
\tilde f\in\spn\{1,\vartheta^1\vartheta^4,\vartheta^2\vartheta^4,\vartheta^3\vartheta^4\}\,,
\eeq
where $\tilde \beta$ is a real parameter. At $\tilde \beta= \theta_0$, we recover \eqref{boundcond123}, while at $\tilde \beta =0$, the supersymmetry is enhanced to $\mathcal{N}=4$ and corresponds to the case described in the next section.
With these boundary conditions, \eqref{btildef} reads
\bal
4 b (\eta_p + \eta_m) =& \left(\Gamma^+ \cos \theta_0 + i \Gamma^- \sin \theta_0 \right) \tilde f (1) 
\\
&+\big(\Gamma^+ \cos \tilde \beta + i \Gamma^- \sin \tilde \beta \big) 
\tilde f (\vartheta^1 \vartheta^4,\vartheta^2 \vartheta^4,\vartheta^3 \vartheta^4) 
\eal
We again find that the space of $1/3$-BPS solutions has two real parameters.

\subsection{$\mathcal{N}=4$}
The boundary conditions given by
\beq
\beta=\theta_0 \left(1-P_\Gamma \right),
\qquad
\gamma=\theta_0 P_1\,,
\qquad
P_\Gamma \tilde f=0\,,
\label{boundcond1234}
\eeq
preserve the supersymmetries generated by $(\vartheta^1, \vartheta^2 , \vartheta^3, \vartheta^4)$, 
and yield a bulk theory conjectured to be dual to the $\mathcal{N}=4$ Chern-Simons quiver 
theory with gauge group $U(N)_k \times U(1)_{-k}$ and a single bi-fundamental hypermultiplet. 

The presence of the projector $P_1$ in the second equation of \eqref{boundcond1234} 
restricts greatly the type of field configurations that we can have. We obtain
\beq
\label{N=4btildef}
4b = \left(\Gamma^+ \cos \theta_0 + i \Gamma^- \sin \theta_0 \right) \tilde f (1) 
+ \Gamma^+ \tilde f' ( \vartheta^1 \vartheta^2, \vartheta^1 \vartheta^3,
 \vartheta^1 \vartheta^4)\, .
\eeq
We can now repeat the procedure of the previous subsections. As in the last two cases, 
the configurations \eqn{n41/2BPS} are excluded by the absence of supersymmetry parameters 
of the form $\Gamma^\pm\vartheta^i$, so we focus again on \eqn{n41/4BPS}. Clearly this 
preserves $\vartheta_1-i\vartheta_2$ and is $1/4$-BPS.

For $\theta_0\neq0$ equation \eqn{N=4btildef} requires $\Gamma^-b$ to be proportional 
to the identity, so in \eqref{n41/4BPS} we need to set $b_{-,1}=b_{-,2}$ 
\eqn{bssbar}. $\Gamma^+b$ is then also related to them. The remaining free parameter is 
$b_{+,1}-b_{+,2}$, so there are overall two parameters for this $1/4$-BPS solution.

A special case is when the lower right block of $b$ vanishes, so 
$b_{-,1}=b_{-,2}=0$. 
This corresponds to case \eqn{n41/2'BPS}, which is compatible with two of our 
preserved supercharges, $\vartheta^1-i\vartheta^2$ and $\vartheta^3-i\vartheta^4$. 
In this case we have only one parameter, since \eqn{N=4btildef} now enforces 
$b_{+,1}=-b_{+,2}$.

We conclude that for these boundary conditions there is a two-parameter family of 
embeddings of the DV-solution which are $1/4$-BPS and a one-parameter family which 
is $1/2$-BPS.

\subsection{$\mathcal{N}=6$}
We will now consider the bulk theory with $n=6$ extended supersymmetry. 
The boundary conditions
\beq
\beta=\theta_0 (1-P_\Gamma)-\theta_0 P_\Gamma\,,
\qquad
\gamma=\theta_0 P_{1,\vartheta^i \vartheta^j}\,,
\qquad
P_{\Gamma, \vartheta^i \vartheta^j \Gamma} \tilde f=0\,,
\label{boundcond123456}
\eeq
where $\vartheta^i\vartheta^j$ stands for all such terms with $i,j=1,\dots,6$, preserve 
the supersymmetries generated by $\vartheta^1, \vartheta^2, \vartheta^3, \vartheta^4,\vartheta^5,\vartheta^6$. Upon adding a $U(M)$ Chan-Paton factor, it is 
proposed in \cite{triality} that the dual theory is the $U(N)_k \times U(M)_{-k}$ ABJ 
model in the large $N,k$, fixed $M$ limit.
We can then proceed as before and write \eqref{btildef} as
\beq
\label{boundscalar6}
b(\eta_p + \eta_m) ={1 \over 4} \left(\Gamma^+ \cos \theta_0  
+ i \Gamma^- \sin\theta_0  \right) \tilde f (1,\vartheta^i \vartheta^j)
\eeq
where we have used the last equation in \eqref{boundcond123456} which tells us 
that $\tilde f$ is only spanned by $1,\vartheta^i \vartheta^j$. 

Now for this $n=6$ case we will work the other way around. Say we want to find 
the configurations which preserve the supersymmetry $\vartheta^1 - i \vartheta^2$.  
Then the most general configuration (that is the one with the greatest number of 
parameters) is given by (as long as $b,\eta_p,\eta_m$ are diagonal)
\beq
\eta_p = \diag (1,0,1,0,1,0,1,0) \, , \quad \eta_m = \diag (0,1,0,1,0,1,0,1)
\eeq
which is thus $1/6$-BPS.
We can then consider degenerate cases by taking some of the non-zero entries 
above to zero. We then find that the following configuration 
\beq
\eta_p = \diag (1,0,0,0,1,0,0,0) \, , \quad \eta_m = \diag (0,1,0,0,0,1,0,0)
\eeq
preserves $\vartheta^1 - i \vartheta^2$ and $\vartheta^3 - i \vartheta^4$, and so 
is $1/3$-BPS. Then the configuration
\beq
\eta_p = \diag (1,0,0,0,0,0,0,0) \, , \quad \eta_m = \diag (0,0,0,0,0,1,0,0)
\eeq
preserves $\vartheta^1 - i \vartheta^2$, $\vartheta^3 - i \vartheta^4$ and  
$\vartheta^5 - i \vartheta^6$, and is thus half-BPS.
However this last case is not compatible with  \eqref{boundscalar6}.

\section{Discussion}

We have developed the tools to study embeddings of solutions of bosonic higher spin 
theory into its supersymmetric extensions and implemented them for the case of the DV 
solution. In the process we also simplified the solution and generalized it to arbitrary 
parity violating phase. The final result of our study is presented in the preceding section, 
where we checked which of the possible embeddings are compatible with the boundary 
conditions and supersymmetries of theories conjectured to be dual to several different 
$3$d Chern-Simons vector models.

One of the theories we considered has an $SU(2)$ flavour symmetry and $\cN=2$ 
supersymmetry. We found $1/2$-BPS solutions parametrized by an arbitrary $2\times2$ 
Hermitian matrix. The matrix structure should be associated to a $U(2)$ flavour symmetry 
(the center of which is normally gauged, in the field theory dual). Thus ignoring the 
center and choosing a Cartan there is a single parameter. 

In the case with $U(1)\times U(1)$ symmetry there are two parameters for the $1/2$-BPS 
solution, which matches with a single free parameter for each element of the Cartan.

This theory can be enhanced (on both sides of the duality) to $\cN=3$. We find though 
the exact same set of BPS solutions, which are now $1/3$-BPS.

Lastly we consider 
a theory with $\cN=4$ supersymmetry. 
Again we find a 
2-parameter family of solutions with the same number of supersymmetries as before, 
which in this case are $1/4$-BPS. 
If we restrict to a subspace of these solutions we find a one dimensional family of 
$1/2$-BPS solutions.

We should mention that part of this analysis was carried out (in a somewhat different 
supersymmetric formalism) in \cite{DV}, and there it was found that the solution preserves 
only $1/4$ of the bulk supersymmetries. We do not understand the reason for the 
discrepancy.

The DV solutions have $SO(3)\times\bR$ global symmetry \cite{DV} so it is
natural to identify them with local operators in the dual field theory and ask 
whether we have found the higher spin dual of all such $1/2$-BPS operators. 

The holographic duality of $4$d higher spin theories has been studied at the level of matching 
of perturbative spectrum and correlation functions. Normally it is very hard to match a 
classical asymptotically $AdS$ bulk solution with a state in the field theory, since the 
spectrum of high-dimension operators in an interacting field theory is very complicated. 
Studying BPS protected operators eliminates this problem, as they can be rather 
easily identified and classified and their 
dimensions are 
fixed by their charges. Our 
analysis allows the first identification of solutions of higher spin theory with states 
in the dual CFT (other than the vacuum).

Indeed, a rudimentary 
examination of the index for some of these theories \cite{index3d} 
indicates 
that the 
dimension of the space of $1/2$-BPS operators does match the Cartan of the flavour 
group, as we have found. More precisely, the three-dimensional index as defined in
 \cite{index3d} is given by a trace formula which gets contributions from states that preserve 
a single supercharge labelled $Q$ (and an associated superconformal generator $Q^\dagger$). 
The trace then contains fugacities that 
correspond to the Cartans of the subalgebra of the full superconformal algebra commuting with $Q,Q^\dagger$,
as well as the Cartans for the flavours. In the spirit of the work done in \cite{Rastelli}, 
the index can be refined further by enhancing supersymmetry, such that the contributions to 
the trace come from states that preserve at least half of the supersymmetries. 
In the three-dimensional case, the only remaining fugacities are then the Cartans of the flavours.  
It would be worthwhile studying this in more detail. Finding 
a precise match should then lead to a quantization of the space of solutions to match 
the discrete dimensions of operators in the dual field theory.

There are obvious generalizations of our analysis. First there are several more 
holographic dualities studied in \cite{triality} and one could implement our results on them. 
In particular the generalization to $n=6$ would allow to study a particular limit of 
ABJ theory. A further generalization, which we have mentioned but have not analysed 
in detail, are theories with extra Chan-Paton factors. These are dual to more general 
Chern-Simons models, whose degrees of freedom are rectangular matrices, not 
just single-column vector models. 

Lastly one can study the embeddings of other classical solutions, like those of 
\cite{firstexact,families,GubserSong}, or find new solutions and embed them.

Supersymmetric higher spin theories provide a fertile ground for deeper understanding 
of the higher spin holographic duality.

\section*{Acknowledgements}

We would like to thank Ofer Aharony, Nicolas Boulanger, Chi-Ming Chang, Slava Didenko, Simone Giombi,
Eric Perlmutter, Mikhail Vasiliev and Cristian Vergu for discussions.
J.B. is grateful for the hospitality of the Yukawa Institute during the course of this work 
and N.D. is grateful for the hospitality of The Weizmann Institute, APCTP, Nordita and CERN. 
The research of J.B. has received funding from the People Programme 
(Marie Curie Actions) of the European 
Union's Seventh Framework Programme FP7/2007-2013/ under REA 
Grant Agreement No 317089 (GATIS). 
The research of N.D. is underwritten by an STFC advanced fellowship 
and the visit to CERN was 
was funded by the National Research Foundation (Korea).

\appendix

\section{Notations and Conventions}
\label{sec:notation}

\subsection{Spinors}
\label{Spinors}
The word \emph{spinor} here designates complex $2$-dimensional vectors 
belonging to the fundamental or anti-fundamental representation of the group 
$Sl(2,\mathbb{C})$ ($SU(2)$) when using a Lorentzian (Euclidian) metric. 
Indices are raised and lowered in the following way
\beq
\psi^{\alpha}=\epsilon^{\alpha\beta}\psi_{\beta}\,,
\qquad
\psi_{\alpha}=\psi^{\beta}\epsilon_{\beta\alpha}, \\
\eeq
where similar rules apply for dotted spinors, and where
\bal
&\epsilon^{12}=-\epsilon^{21}=\epsilon_{12}=-\epsilon_{21}=1 \\
&\epsilon_{\alpha\beta}\epsilon^{\beta \gamma}=-\delta_\alpha^\gamma \\
&\epsilon^{\alpha\beta}=\bar{\epsilon}^{\dot \alpha \dot \beta}
\eal
In Lorentzian metric, we choose the explicit form for the hermitian soldering form
\beq
\sigma^0={I \over \sqrt 2}\,,
\qquad
\sigma^1={1 \over \sqrt 2} \begin{pmatrix}
0& 1 \\ 1& 0
\end{pmatrix},
\qquad
\sigma^2={1 \over \sqrt 2} \begin{pmatrix}
0&-i \\ i& 0
\end{pmatrix},
\qquad\sigma^3={1 \over \sqrt 2} \begin{pmatrix}
1& 0 \\ 0&-1 \end{pmatrix},
\eeq
so that we have the following useful relations
\bal
&(\bar{\sigma}^a)^{\dot \alpha \beta} 
\equiv \epsilon^{\dot \alpha \dot \gamma} 
\epsilon^{\beta \delta} (\sigma^a)_{\delta \dot \gamma}\,,
\quad&&
\Tr [\sigma^a \bar \sigma^b ]=\eta^{ab}\,,\\
& (\sigma^a)_{\alpha \dot \beta} (\bar{\sigma}_a)^{\dot \gamma \delta} 
=\delta_\alpha^\delta \delta_{\dot \beta}^{\dot \gamma}\,,
\quad&&
(\sigma^a)_{\alpha \dot \beta} (\sigma_a)_{\gamma \dot \delta} 
=\epsilon_{\alpha \gamma} \epsilon_{\dot \beta \dot \delta}\,,
\\
& (\sigma^a \bar \sigma^b+\sigma^b \bar \sigma^a)_\alpha^{\ \beta} 
=\eta^{ab} \delta_\alpha^{\beta}\,.
\eal
The maps from multispinors to tensors of $SO(3,1)$ are then given by
\begin{align}
T_{\mu_1 \dots \mu_n}^{\nu_1 \dots \nu_s}
&=(\sigma^{\nu_1})_{\alpha_1 \dot{\alpha}_1} 
 \cdots (\sigma^{\nu_s})_{\alpha_s \dot{\alpha}_s} 
 (\sigma_{\mu_1})^{\dot \beta_1 \beta_1} 
 \cdots (\sigma_{\mu_n})^{\dot \beta_n \beta_n} 
 T_{\beta_1 \dot \beta_1 \dots \beta_n \dot \beta_n}^{\dot \alpha_1 \alpha_1 \dots \dot \alpha_s \alpha_s}\,,
\\
T_{\beta_1 \dot \beta_1 \dots \beta_n \dot \beta_n}^{\dot \alpha_1 \alpha_1 \dots \dot \alpha_s \alpha_s} 
&=(\sigma_{\nu_1})^{\dot{\alpha}_1 \alpha_1} 
\cdots (\sigma_{\nu_s})^{\dot{\alpha}_s \alpha_s} 
(\sigma^{\mu_1})_{\beta_1 \dot \beta_1} 
\cdots (\sigma^{\mu_n})_{\beta_n \dot \beta_n} T_{\mu_1 \dots \mu_n}^{\nu_1 \dots \nu_s}.
\end{align}
The ones we will be particularly interested in are
\bal
\label{h-omega}
h^a&=(\bar\sigma^a)^{\dot \gamma \alpha} h_{\alpha \dot \gamma}\,,
&\quad
\omega_{ab}
&=-i \left((\sigma_{ab})^{\dot \beta}_{\ \dot \alpha} \omega^{\dot \alpha}_{\ \dot \beta}
-(\sigma_{ab})_{\alpha}^{\ \beta} \omega^{\ \alpha}_{\beta} \right)\,,
\\
h_{\alpha\dot \beta}&=h^a (\sigma_a)_{\alpha \dot \beta}\,,
&\quad
\omega_\alpha^{\ \beta}&=-{i \over 2} \omega_{a b} (\sigma^{ab})_\alpha^{\ \beta}\,,
\qquad
\omega^{\dot \alpha}_{\ \dot \beta}={i \over 2} \omega_{ab} (\bar \sigma^{ab})^{\dot \alpha}_{\ \dot \beta},
\eal
where we have defined the following hermitian matrices
\beq
(\sigma^{ab})_\alpha^{\ \beta} 
\equiv \frac{i}{2} (\sigma^a \bar \sigma^b-\sigma^b \bar \sigma^a)_\alpha^{\ \beta},
\qquad
(\bar \sigma^{ab})_{\ \dot \alpha}^{\dot \beta} 
\equiv \frac{i}{2} (\bar \sigma^a \sigma^b-\bar \sigma^b \sigma^a)_{\ \dot \alpha}^{\dot \beta},
\eeq
which satisfy the Lorentz algebra
\beq
\left[\sigma^{ab} , \sigma^{cd} \right]
=i \left(\eta^{bc} \sigma^{ad}-\eta^{ac} \sigma^{bd}+\eta^{ad} \sigma^{bc}-\eta^{bd} \sigma^{ac} \right).
\eeq
as well as
\beq
(\sigma_{a b})_{\alpha}^{\ \beta} (\sigma^{a b})_{\gamma}^{\ \delta} 
=\delta_{\alpha}^{\delta} \delta_{\gamma}^{\beta} 
-\epsilon_{\alpha \gamma}\epsilon^{\beta\delta}\,,
\qquad
(\bar\sigma_{a b})_{\ \dot \alpha}^{\dot \beta} (\sigma^{a b})_{\gamma}^{\ \delta}=0.
\eeq

We can then define the four dimensional $\gamma$ matrices
\beq
\gamma^a=\begin{pmatrix}
0& \sigma^a \\ \bar \sigma^a& 0 \\
\end{pmatrix},
\qquad
\gamma^{ab} \equiv \frac{i}{2} \left[\gamma^a , \gamma^b \right]=\begin{pmatrix}
\sigma^{ab}& 0 \\ 0& \bar \sigma^{ab}
\end{pmatrix}.
\eeq
with
\bal
\left\{ \gamma^a , \gamma^b \right\} = \eta^{a b}.
\eal

\subsection{Star-Product Formulae}
\label{sec:star}

The equations of motion \eqref{masterfull21}-\eqref{masterfull22} are written in this compact 
form with the use of a star-product defined as
\bal
\Phi(Y,Z) \star \Theta(Y,Z)&=\Phi(Y,Z) \exp \left[-\epsilon^{\alpha\beta} 
\big({\mathop\partial^\leftarrow}_{y^\alpha}+{\mathop\partial^\leftarrow}_{z^\alpha} \big) 
\big({\mathop\partial^\rightarrow}_{y^\beta}-{\mathop\partial^\rightarrow}_{z^\beta} \big)+\text{c.c.} \right] \Theta(Y,Z)\\
&=\int d^2 u d^2 \bar{u} d^2v d^2 \bar{v}\, \Phi(Y+U,Z+U) \Theta(Y+V,Z-V) 
e^{\left(u_\alpha v^\alpha+\bar{u}_{\dot \alpha} \bar{v}^{\dot \alpha} \right)}
\eal
where the integral in the second line is implicitly normalized and the integration contour 
chosen in such a way that
\begin{equation}
\Phi \star 1=1 \star \Phi=1.
\end{equation}
More precisely, this means that we assume the following representation of the delta function
\beq
\int du^2 dv^2 e^{u_\alpha v^\alpha}=\int du^2 \delta^2(u)\,,
\eeq
and similarly for the dotted variables. This is equivalent to integrating over the real axis 
rotated by a certain angle. We see that if $u_1$ and $v_2$ are rotated by $\pi/4$, then 
we obtain a valid representation of the delta function. However, the rotation angle is not 
entirely fixed, as can be seen from the fact that we could have chosen $u_1$ and $v_2$ 
to be both rotated by $-\pi/4$ instead. 
Our choice will be the following
\bal
& u_1 \, , \, v_2 \quad \text{rotated by } -\frac{\pi}{4} \\
& u_2 \, , \, v_1 \quad \text{rotated by } \quad \, \frac{\pi}{4}
\eal
and similarly for the dotted variables. 

With this definition, we have
\begin{align}
[y_\alpha , y_\beta]_\star&=-2 \epsilon_{\alpha \beta}\,,\qquad
&[\bar{y}_{\dot \alpha} , \bar{y}_{\dot \beta}]_\star&=-2 \epsilon_{\dot \alpha \dot \beta}\,,\qquad
&[y_\alpha , \bar{y}_{\dot \beta}]_\star&=0 \,,\\
[z_\alpha , z_\beta]_\star&=2 \epsilon_{\alpha \beta}\,,\qquad
&[\bar{z}_{\dot \alpha} , \bar{z}_{\dot \beta}]_\star&=2 \epsilon_{\dot \alpha \dot \beta}\,,\qquad
&[z_\alpha , \bar{z}_{\dot \beta}]_\star&=0 \, ,
\end{align}
as well as the following properties which will be used in Appendix~\ref{sec:DV}
\beq
v = \delta(y) \star \delta(z) \, , \qquad \delta(y) \star \delta(y) = \delta(z) \star \delta(z) =1.
\eeq
Note that the bilinears
\begin{equation}
L_{\alpha \beta}=\half \left\{y_\alpha, y_\beta \right\}\,,
\qquad
\bar{L}_{\dot \alpha \dot \beta}=\half \{\bar{y}_{\dot \alpha}, \bar{y}_{\dot \beta} \}\,,
\qquad
P_{\alpha \dot \beta}=y_\alpha \bar{y}_{\dot \beta}\,,
\end{equation}
give an oscillator representation of the algebra $so(3,2) \simeq sp(4)$, which is the unique 
finite dimensional subalgebra of the higher spin algebra spanned by general polynomials in $Y$ of the form
\begin{equation}
\label{general pol}
P(y,\bar{y})=\sum_{n,m=0}^\infty P^{\alpha_1 \dots \alpha_n, \dot{\alpha}_1 \dots \dot{\alpha}_m} y_{\alpha_1} \dots y_{\alpha_n} \bar{y}_{\dot{\alpha}_1} \dots \bar{y}_{\dot{\alpha}_m}.
\end{equation}
In perturbation theory, it is often needed to evaluate the star product of 
$W_0$ \eqn{vac1} with an arbitrary function. If $\Phi$ is a zero-form, we obtain
\begin{align}
\left[W_{\alpha \beta} y^\alpha y^\beta , \Phi(Y,Z) \right]_\star 
&=-4 W_{\alpha \beta} \epsilon^{\alpha \gamma} \epsilon^{\beta \delta}
\left\{y_\delta \partial_{y^\gamma}
+\partial_{y^{\gamma}} \partial_{z^{\delta}} \right\} \Phi(Y,Z) \\
\left[W_{\dot \alpha \dot \beta} \bar{y}^{\dot \alpha} \bar{y}^{\dot \beta} , \Phi(Y,Z) \right]_\star 
&=-4W_{\dot \alpha \dot \beta}\epsilon^{\dot \alpha \dot \gamma} \epsilon^{\dot \beta \dot \delta} 
\left\{\bar{y}_{\dot\delta} \partial_{\bar{y}^{\dot \gamma}}
+\partial_{\bar{y}^{\dot \gamma}} \partial_{\bar{z}^{\dot \delta}} \right\} \Phi(Y,Z) \\
\left[W_{\alpha \dot \beta} y^\alpha \bar{y}^{\dot \beta} , \Phi(Y,Z) \right]_\star 
&=-2W_{\alpha \dot \beta}\epsilon^{\alpha \gamma}\epsilon^{\dot \beta \dot \delta}
\left\{\bar{y}_{\dot\delta}\partial_{y^\gamma}
+y_\gamma\partial_{\bar y^{\dot \delta}} 
+\left(\partial_{z^\gamma} \partial_{\bar y^{\dot \delta}}+\partial_{y^\gamma} \partial_{\bar{z}^{\dot \delta}} \right) \right\}\Phi(Y,Z) 
\end{align}
If $\Phi$ is a one-form, it will appear in the equations in anti-commutators, and we can use the 
above formula without changing signs, and replacing 
$\left[\cdot, \cdot \right]_\star \rightarrow \left\{\cdot, \cdot \right\}_\star$.
If $\Phi$ transforms in the so-called twisted adjoint representation, 
we are led to evaluate quantities of the form $\Phi \star \pi \left(W_0 \right)-W_0 \star \Phi$. 
We obtain
\bal
&\Phi \star \pi \left(W_0 \right) -W_0 \star \Phi 
=-\left[W_{\alpha \beta} y^\alpha y^\beta , \Phi(y,\bar y, z) \right]_\star 
-\left[W_{\dot \alpha \dot \beta} \bar{y}^{\dot \alpha} \bar{y}^{\dot \beta} , \Phi(y,\bar y, z) \right]_\star
\\
&\quad{}+W_{\alpha \dot \beta} \left[-2 y^\alpha \bar{y}^{\dot \beta} 
-2 \epsilon^{\alpha \gamma} \bar{y}^{\dot \beta} \partial_{z^\gamma} 
-2\epsilon^{\dot \beta \dot \gamma} y^\alpha \partial_{\bar z ^{\dot \gamma}} 
-2 \epsilon^{\alpha \gamma} \epsilon^{\dot \beta \dot \delta}
 \left( \partial_{y^\gamma} \partial_{\bar{y}^{\dot \delta}} 
 +\partial_{z^\gamma} \partial_{\bar{z}^{\dot \delta}} \right) \right] \Phi(Y,Z).
\label{twistedadjoint}
\eal

\subsection{Conventions in the Literature}
\label{sec:conventions}

We will here briefly describe different conventions used by other authors. Setting aside 
conventions used for spinors and soldering forms, the most fundamental choice of 
convention is usually made at the level of the definition of the canonical commutation 
relations of the auxiliary spinor variables. We have opted for
\beq
\left[y_\alpha , y_\beta \right]_\star=-\left[z_\alpha , z_\beta \right]_\star=-2 \epsilon_{\alpha \beta}\,,
\qquad
\left[\bar y_{\dot \alpha} , \bar{y}_{\dot \beta} \right]_\star
=-\left[\bar{z}_{\dot \alpha} , \bar{z}_{\dot \beta} \right]_\star=-2 \epsilon_{\dot \alpha \dot \beta}.
\eeq
In many places, one will find the following definitions
\beq
\left[y_\alpha , y_\beta \right]_\star=-\left[z_\alpha , z_\beta \right]_\star=2i \epsilon_{\alpha \beta}\,,
\qquad
\left[\bar y_{\dot \alpha} , \bar{y}_{\dot \beta} \right]_\star
=-\left[\bar{z}_{\dot \alpha} , \bar{z}_{\dot \beta} \right]_\star=2i \epsilon_{\dot \alpha \dot \beta}.
\eeq
With this choice, the star-product \eqref{starproduct} becomes
\beq
f(Y,Z) \star g(Y,Z)
=f(Y,Z) \exp i\left[\epsilon^{\alpha\beta} 
\left({\mathop\partial^\leftarrow}_{y^\alpha}+{\mathop\partial^\leftarrow}_{z^\alpha}\right)
\left({\mathop\partial^\rightarrow}_{y^\beta}-{\mathop\partial^\rightarrow}_{z^\beta}\right)
+\text{c.c.}\right] g(Y,Z).
\eeq
Hermitian conjugation can of course still be defined, but in a less trivial way. Looking 
again at the simplest case of the $AdS$ vacuum solution, we see that the definition 
\eqref{vac1}-\eqref{vac2} is no longer valid, in the sense that the components do no 
longer correspond to the ones that satisfy the first order formulation of the Einstein equations. 
To circumvent this, 
authors usually expand the field $W$ as (see \cite{99review})
\beq
W(x,y,\bar y)=W_\mu (x,y, \bar y) dx^\mu=
\!\!\!\!\!\!\sum_{\substack{n,m=0\\n+m=even}}^{\infty}\!\!\! {1 \over 2 i n! m!}
W_{\alpha_{1} \dots \alpha_{n}, \dot{\alpha}_1 \dots \dot{\alpha}_m} 
y^{\alpha_1} \cdots y^{\alpha_n} \bar{y}^{\dot{\alpha}_1} \cdots \bar{y}^{\dot{\alpha}_m}\,,
\eeq
where the important point is the presence of the factor $i$. Then one can identify 
the components of the bilinears in the oscillators with the Lorentz connection and 
vielbein as we did, as long as the reality condition is chosen to be 
$(W)^\dagger=-W$. For the fully non-linear theory, with the $Z$ dependence 
included, one has to modify the way the oscillators transform
\beq
(y^\alpha)^\dagger=\bar y^{\dot \alpha}\,,
\qquad(z^\alpha)^\dagger=-\bar z^{\dot \alpha}\,,
\eeq
so that for two functions $f,g$ of $(Y,Z|x)$ we then obtain
\beq
\left(f \star g \right)^\dagger=g^\dagger \star f^\dagger\,,
\eeq
where it should be noted that the order of the functions has been inverted. Once more 
one has to introduce appropriate minus signs when manipulating forms, as the hermitian 
conjugation is blind to the wedge product.
For consistency with the equations of motion involving the field $B$, this in turn imposes 
that $B^\dagger=v \star B \star v$.
Finally, we would like to point out some possible differences in the definitions of the 
master field $S$. Indeed, $S_0$ can be taken to be identically zero, in which case 
the differential operator in \eqref{masterfull21} should be replaced as follows
\begin{equation}
d=dx^\mu {\partial \over \partial x^\mu}
\quad\rightarrow\quad
\hat d \equiv dx^\mu {\partial \over \partial x^\mu}+dz^{\alpha} {\partial \over \partial z^{\alpha}}
+d\bar{z}^{\dot \alpha} {\partial \over \partial \bar{z}^{\dot \alpha}}.
\end{equation}

\section{The Didenko-Vasiliev solution}
\label{sec:DV}

We review here the construction of the Didenko-Vasiliev solution \cite{DV}. 
We start with the $AdS_4$ vacuum solution, then the extremal $AdS$-black-hole 
solution. Then we present the solution of the free higher-spin theory and finally the full 
solution to the interacting theory employed in the main text.

\subsection{The Vacuum Solution}
The $AdS_4$ vacuum solution in the interacting theory is given by
\begin{align}
W_0&=-\frac{1}{4} \left(\omega_{\alpha \beta} y^\alpha y^\beta
+\omega_{\dot \alpha \dot \beta} \bar{y}^{\dot \alpha} \bar{y}^{\dot \beta} 
-\sqrt 2 h_{\alpha \dot \beta} y^\alpha \bar{y}^{\dot \beta} \right), \label{vac1} \\
B_0&=0\,,
\qquad
S_0=z_\alpha dz^\alpha 
+\bar{z}_{\dot \alpha} d \bar{z}^{\dot \alpha}. \label{vac2}
\end{align}
One can define a perturbation scheme around the vacuum solution, and the obtained 
linearised equations of motion are equivalent to the ones written by Fronsdal in~\cite{Fronsdal} 
for real massless fields propagating in $AdS_4$ (see \cite{99review} for an explicit derivation).

In this paper we will be using global coordinates for $AdS_4$
\beq
ds^2=g^0_{\mu\nu}dx^\mu dx^\nu
\equiv(1+\lambda^{-2}r^2)dt^2-\frac{1}{1+\lambda^{-2}r^2}dr^2
-r^2(d\theta^2+\sin^2\theta\,d\varphi^2)\,.
\label{Adsmetric}
\eeq
From here onwards we set the $AdS$ scale $\lambda \rightarrow 1$ for simplicity. 
We choose the following explicit expressions for the vielbeins
\beq
h^0=\sqrt{1+r^2} \,dt\,,
\qquad
h^1=\frac{1}{\sqrt{1+r^2}}\,dr\,,
\qquad
h^2=r \,d\theta\,,
\qquad
h^3=r \sin \theta \,d \varphi\,,
\eeq
and obtain the connection one-forms
\beq
\omega_{01}=r\,dt\,,
\qquad
\omega_{12}=\sqrt{1+r^2}\,d \theta\,,
\qquad
\omega_{13}=\sqrt{1+r^2}\sin \theta\,d \varphi\,,
\qquad
\omega_{23}=\cos \theta\,d \varphi\,,
\eeq
with all others are zero. The connections and vielbein 
$\omega_{\alpha\beta}$, $\omega_{\dot \alpha \dot \beta}$, $h_{\alpha \dot \beta}$ 
appearing in \eqref{vac1} are then obtained using Appendix~\ref{Spinors}.

\subsection{Black-Holes in $AdS_4$}
The simplest black holes in $4$-dimensions are characterized by an $SO(3) \times \mathbb R$ 
symmetry. The black-hole solution in $AdS_4$ can be written in the so-called Kerr-Schild form 
(see~\cite{Carter} and~\cite{Gibbons})
\beq
g_{\mu\nu}=g^0_{\mu\nu}-\frac{2M}{r}k_\mu k_\nu\,,
\qquad
g^{\mu\nu}=g^{0\mu\nu}+\frac{2M}{r}k^\mu k^\nu\,,
\eeq
with
\beq
k_\mu dx^\mu=dt-\frac{dr}{1+ r^2}.
\label{kerrschildvec}
\eeq

Using this definition, one can define the following traceless completely symmetric tensors
\beq
\phi_{\mu_1 \dots \mu_s}=\frac{2M}{r} k_{\mu_1} \dots k_{\mu_s}\,,
\label{tensors}
\eeq
which satisfy the equations of motion of a massless spin-$s$ field in a $AdS$ background
\beq
D^\mu D_\mu \phi_{\nu(s)}-s D_\mu D_\nu \phi^\mu_{\nu(s-1)}
=-2(s-1)(s+1)\phi_{\nu(s)}.
\eeq

From these facts we learn a number of things which will be directly transposed 
in the construction of the fully interacting theory. First, a black-hole solution in 
$AdS_4$ can be written as a one loop perturbation. The perturbation is 
constructed from a vector $k$ which in turn generates an infinite tower 
of massless higher-spin fields. Finally, it should be noted that this Kerr 
vector $k$ can be expressed in terms of the killing vector 
$V = \sqrt{2} \partial/\partial t$ and the associated Killing 
two-form $\kappa_{\alpha \beta},\kappa_{\dot \alpha \dot \beta}$ through this simple formula
\beq
k_{\alpha \dot \alpha} = {1 \over 1 +r^2} 
\left( v_{\alpha \dot \alpha} - {\kappa_\alpha{}^{\beta} v_{\beta \dot \alpha} \over r} \right).
\eeq
In the next section we explain in detail the role of these quantities.

\subsection{Killing Matrix}
\label{sec:symmetries}

The main ingredients in the construction of the Didenko-Vasiliev (DV) solution are the 
Killing vectors and Killing two-forms of the vacuum. From \eqref{gauge}, we see that 
a global symmetry is generated by a $Z$-independent gauge parameter 
$\epsilon=\epsilon(Y|x)$ satisfying
\beq
d\epsilon-[W_0,\epsilon]_\star=0\,.
\label{global}
\eeq
If we take $\epsilon$ to be bilinear in $Y$ and identify $v_{\alpha\dot\beta}$ with a Killing vector 
$V_\mu$, and $\kappa_{\alpha \beta} ,\, \kappa_{\dot \alpha \dot \beta}$ with the self and anti-self 
dual parts of the Killing two-form $V_{\mu \nu}=D_{[\mu} V_{\nu]}$ associated with the 
same Killing vector, \eqref{global} reduces to the Killing equation with the identification
\begin{equation}
\label{KillingDV}
K_{AB}=\begin{pmatrix}
 \sqrt{2} \kappa_{\alpha \beta}& v_{\alpha \dot \beta} \\ v_{\alpha \dot \beta}&  \sqrt{2} \kappa_{\dot \alpha \dot \beta}
\end{pmatrix}\, ,
\end{equation} 
to which we will be referring to as the Killing matrix. In $Sp(4)$ notations, 
\eqref{global} can be written in a compact way as
\begin{equation}
D_0 \left(K_{AB} Y^AY^B\right)=0.
\end{equation}

In what follows we will be using a particular time-like Killing vector defined as
\beq
\label{Vdt}
V=V^\mu \frac{\partial}{\partial x^\mu}=\sqrt 2 \frac{\partial}{\partial t}
\eeq
The various factors in the definition of the Killing matrix in terms of the Killing vector and Killing 
two-forms have been chosen so that we have the following identity
\beq
K_A^{\ B} K_{B}^{\ C}=-\delta_A^C\,,
\eeq
which reads in components
\begin{align}
&\kappa_{\alpha}^{\ \beta} \kappa_{\beta \gamma} \equiv-\epsilon_{\alpha \gamma} \kappa^2\,,
&
&\kappa^2 \equiv \half \kappa_{\alpha \beta} \kappa^{\alpha \beta}=\det(\kappa_{\alpha\beta})\,,
\label{killingeq1} \\
& 2  \kappa_\alpha^{\ \beta} \kappa_{\beta \gamma}
+v_\alpha^{\ \dot \beta} v_{\dot \beta \gamma}=-\epsilon_{\alpha \gamma}\,,
&
&2  \kappa_{\dot \alpha}^{\ \dot \beta} \kappa_{\dot \beta \dot \gamma}+v_{\dot \alpha}^{\ \beta} v_{\beta \dot \gamma}=-\epsilon_{\dot \alpha \dot \gamma}\,,
\label{killingeq2} \\
&\kappa_{\alpha}^{\ \beta}v_{\beta \dot \gamma}+v_\alpha^{\ \dot \beta} \kappa_{\dot \beta \dot \gamma}=0\,,
&
&\kappa_{\dot \alpha}^{\ \dot \beta} v_{\dot \beta \gamma}+v_{\dot \alpha}^{\ \beta}\kappa_{\beta \gamma}=0. \label{killingeq3}
\end{align}
If we contract the second set of these equations, we obtain
\beq
2  \kappa^2+v^2= 2  \bar{\kappa}^2+v^2=1\,,
\label{contract}
\eeq
where we have defined $v^2 \equiv \half v_{\alpha \dot \beta} v^{\alpha \dot \beta}=\half V_\mu V^\mu$ 
which is consistent with $\kappa^2=\bar{\kappa}^2$.
More explicitly, we have
\beq
v^2=1+ r^2\,,
\qquad
\kappa^2=-{r^2 \over 2} .
\eeq
We will also be needing explicit expressions for $\kappa_{\alpha\beta}$
\bal
\kappa_{\alpha\beta}
&= { \sqrt{2} \over 2} r \begin{pmatrix}-1& 0 \\ 0& 1 \end{pmatrix}_{\alpha\beta}\,,
&\qquad
\kappa_{\dot \alpha \dot \beta}
&= { \sqrt{2} \over 2} r \begin{pmatrix}-1& 0 \\ 0& 1 \end{pmatrix}_{\dot \alpha \dot \beta}\,,
\\
v_{\alpha \dot \beta}
&=\sqrt{1+ r^2} \begin{pmatrix} 1& 0 \\ 0& 1 \end{pmatrix}_{\alpha \dot \beta}. 
\label{explicitspin}
\eal
Of course, it is straightforward to recover \eqref{killingeq1}-\eqref{contract} from \eqref{explicitspin}.

The linearised equations of motion are given by
\begin{align}
D_0 \Omega 
&\equiv d \Omega-W_0 \wedge_\star \Omega-\Omega \wedge_\star W_0 
\nn \\
&=h^{\gamma \dot \alpha} \wedge 
h_\gamma^{\ \ \dot \beta} \partial_{\dot \alpha} \partial_{\dot \beta} C(0,\bar y|x) 
+h^{\alpha \dot \gamma} \wedge h^{\beta}_{\ \dot \gamma} \partial_\alpha \partial_\beta C(y,0|x)\,,
\label{master1} \\
\tilde D_0 C
&\equiv dC-W_0 \star C+C \star \pi(W_0)=0\,,
\label{master2}
\end{align}
where the master fields $\Omega,C$ correspond to the $Z$-independent part of $(W,B)$. 
They are solved by the following ansatz
\beq
C=bF_K \star \delta(y)\,,\label{ansatzC}
\eeq
with
\beq
F_K \equiv 4 \exp \left(\frac{i}{2} K_{AB} Y^AY^B \right). \label{FK}
\eeq
This solution has very interesting properties
\beq
F_K \star \delta(y)=F_K \star \delta(\bar y)\,,
\qquad
F_K \star F_K=F_K.
\eeq
By performing the star-product explicitly, we obtain
\begin{equation}
C=\frac{4b}{ r} 
\exp\left[\frac{i}{2  \kappa^2} \left(\kappa_{\alpha \beta}y^\alpha y^\beta
+\kappa_{\dot \alpha \dot \beta} \bar{y}^{\dot \alpha} \bar{y}^{\dot \beta}
+2i \kappa_{\alpha \gamma}v^\gamma_{\ \dot \beta} y^\alpha \bar{y}^{\dot \beta} \right) \right].
\end{equation}
We can read off the components of the generalised higher-spin Weyl tensors 
\beq
C_{\alpha(2s)}=\frac{b}{s! 2^{s-2}  r} 
\left(\frac{i \kappa_{\alpha\alpha}}{ \kappa^2} \right)^s\,,
\qquad
\bar C_{\dot \alpha (2s)}=\frac{b}{s! 2^{s-2}  r} 
\left(\frac{i \kappa_{\dot \alpha \dot \alpha}}{ \kappa^2} \right)^s. 
\label{petrovWeyl}
\eeq
This corresponds at the spin two level to a Petrov type-D Weyl tensor, describing 
a black hole.

\subsection{Didenko-Vasiliev Solution for the Purely Bosonic Theory}

The solution to the non-linear equations \eqref{masterfull1}-\eqref{masterfull4} was 
obtained in~\cite{DV} by starting from the solution of the linearised theory 
and studying the corrections order by order in $b$. It turns out that in a similar way 
to what happens for the Kerr-Schild ansatz for classical GR, there are no higher 
order corrections beyond the linearised part. 

\subsubsection{Induced Star-Product and Fock Space}
To adapt the solution of the free theory to the interacting one, the idea is to use 
an ansatz for the linearised part of the master fields of the form $F_K \star f(Z|x)$ 
where the $Z$-dependent part is explicitly factored out. If one defines 
$A_A$ by
\beq
F_K\star Z_A\equiv F_K\,A_A\,,
\qquad
A_A\equiv(a_\alpha, \bar a_{\dot \alpha})\equiv Z_A+iK_A^{\ B}Y_B\,,
\qquad
\left[A_A,A_B\right]=4 \epsilon_{AB}.
\eeq
then in fact the star product of $F_K$ with any function of $Z$ can be expressed without 
the star product in terms of $A$
\beq
F_K\star\phi(Z|x)=F_K\,\phi(A|x)\,.
\eeq
Furthermore, it can be shown that functions of this form define a subalgebra as
\beq
\left(F_K \phi_1 \right) \star \left(F_K \phi_2 \right)=F_K \left(\phi_1 * \phi_2 \right),
\label{subalgebra}
\eeq
where the star-product $*$ is associative and takes the form
\beq
(\phi_1*\phi_2) (A)=\int d^4u\, \phi_1 (A+2U_+) \phi_2(A-2U_-)e^{2U_{+A} U^A_{-}}. \label{induced}
\eeq
where we have defined $U_{\pm A}=\Pi_{\pm A}^{\ \ B} U_B$ in terms of the projectors
\beq
\label{ProjectorsPi}
\Pi_{\pm AB}=\half \left(\epsilon_{AB} \pm iK_{AB} \right)\,,
\qquad
\Pi_{\pm A}^{\ \ B}\Pi_{\pm B}^{\ \ C}=\Pi_{\pm A}^{\ \ C}\,,
\qquad
\Pi_{\pm A}^{\ \ B}\Pi_{\mp B}^{\ \ C}=0 .
\eeq
We can then obtain the following formulae for any function $\phi(a)$ holomorphic in $a$
\beq
\left[a_\alpha, \phi(a) \right]_*=2 \partial_{a^\alpha}\phi(a)\,,
\qquad
\left\{a_\alpha, \phi(a) \right\}_* 
=2 \left(a_\alpha+i  \kappa_\alpha^{\ \beta}\partial_{a^\beta} \right) \phi(a)\,,
\label{inducedcommutation}
\eeq
where similar equations hold for functions that depend only on $\bar a$.
The star-product \eqref{induced} possesses Kleinien operators $\mathcal K$ and $\bar{\mathcal{K}}$, defined as follows
\beq
F_K \star \delta(z)=F_K \mathcal K\,,
\qquad
F_K \star \delta(\bar z)=F_K \bar{\mathcal{K}}.
\eeq
Their explicit expressions are
\beq
\mathcal K
=\frac{1}{r}\exp\left[\frac{i\kappa_{\alpha\beta}}{2\kappa^2}a^\alpha a^\beta\right],
\qquad
\bar{\mathcal{K}}=\frac{1}{r}\exp\left[\frac{i\kappa_{\dot\alpha\dot\beta}}{2\kappa^2}
\bar{a}^{\dot\alpha}\bar{a}^{\dot\beta}\right],
\label{inducedK}
\eeq
and it is then straightforward to show that
\bal
&\mathcal{K} * \mathcal{K}=\bar{\mathcal{K}} *\bar{\mathcal{K}}=1\,,
\qquad
\left\{\mathcal K , a_\alpha \right\}_*=\left\{\bar{\mathcal{K}} , \bar{a}_{\dot \alpha} \right\}_*=0\,,
\\
& \left[\mathcal K ,\bar{\mathcal{K}} \right]_*=\left[\mathcal K , \bar{a}_{\dot \alpha} \right]_*=\left[\bar{\mathcal{K}} , a_\alpha \right]_*=0. \label{Kcommute}
\eal

As for the action of $F_K$ on the $Y$ coordinates, it is easy to see that it annihilates half of them
\beq
\label{leftright}
Y_{-A}\star F_K=F_K \star Y_{+A}=0\,,
\eeq
Using the closure \eqn{subalgebra}, we can immediately see that 
this projection holds true for any member of the subalgebra
\beq
\label{generalprojphi}
(F_K\,\phi)\star Y_{+A}=(F_K\,\phi)\star F_K\star Y_{+A}=0.
\eeq

\subsubsection{Solving The Non-Linear Equations Of Motion}

We have thus seen that by choosing to factor out the $Z$-dependence by looking at 
functions of the form $F_K\star\phi(Z|x)$ we are led to study a subalgebra of the full 
higher-spin algebra. This can be used to simplify the equations of motions. More 
precisely, we take the following ansatz for the master fields of the theory
\beq
W=W_0+F_K \left[\Omega(a|x)+\bar{\Omega}(\bar{a}|x)\right].
\eeq
\beq
S_\alpha=z_\alpha+F_K \sigma_\alpha(a|x)\,,
\qquad
\bar{S}_{\dot \alpha}=\bar{z}_{\dot \alpha}+F_K \bar{\sigma}_{\dot \alpha}(\bar{a}|x)\,,
\label{finalansatz2}
\eeq
while the $B$-field is kept identical to \eqref{ansatzC}
\beq
B=b F_K \star \delta(y). \label{finalansatz1}
\eeq
This ansatz reduces the equations \eqref{masterfull5} and \eqref{masterfull4} 
to the following set of equations
\bal
&\left[\varsigma_\alpha(a|x), \varsigma_\beta(a|x) \right]_*
=2\epsilon_{\alpha \beta} \left(1+e^{i \theta} b \mathcal{K} \right),
&\qquad
&\left[\varsigma_\alpha(a|x),
\bar\varsigma_{\dot\alpha}(\bar a|x) \right]_*=0\,,
\\
&\left\{\mathcal{K}, \varsigma_\alpha(a|x) \right\}_*=0\,,
&\qquad
&\left\{\varsigma_\alpha(a|x),\bar{\mathcal{K}}\right\}_*=0\,,
\\
&\mathcal{Q} \Omega-\Omega \wedge_* \Omega=0\,,
&\qquad
&\mathcal{Q} \varsigma_\alpha-\left[\Omega, \varsigma_\alpha \right]_*=0\,,
\label{wigner}
\eal
with similar equations in the barred sector and where we have defined
\begin{align}
\varsigma_\alpha&=a_\alpha+\sigma_\alpha(a|x)\,,
\qquad
\bar{\varsigma}_{\dot \alpha}=\bar{a}_{\dot \alpha}+\bar{\sigma}_{\dot \alpha}(\bar{a}|x)\,,\\
\mathcal{Q}
&=\left(\hat{d}-{i \over 2}  d\kappa^{\alpha \beta}\partial_{a^\alpha}\partial_{a^\beta} \right) \, ,
\end{align}
where $\hat d$ acts as the standard exterior derivative but leaves invariant $a,\bar a$ 
{\em i.e.}, $\hat d a = \hat d \bar a =0$. The simplified set of equations \eqref{wigner} can be solved 
exactly by analogy with the standard perturbative analysis, taking here $b$ to be the infinitesimal 
parameter. We obtain
\begin{align}
\sigma_\alpha(a|x)
&=\frac{be^{i\theta}}{r} \pi_\alpha^{+\beta}a_\beta
\int_0^1 dt \exp \left({it \over 2} {\kappa_{\alpha \beta} \over  \kappa^2} a^\alpha a^\beta\right),
\label{sigma} \\
\bar{\sigma}_{\dot \alpha}(\bar{a}|x)
&= \frac{be^{-i\theta}}{r} \pi_{\dot \alpha}^{+\dot \beta}\bar{a}_{\dot \beta}
\int_0^1 dt \exp \left({it \over 2} {\kappa_{\dot \alpha \dot \beta} \over \kappa^2} \bar{a}^{\dot \alpha} \bar{a}^{\dot \beta} \right), \label{sigmabar}
\end{align}
with a new set of projectors 
\beq
\pi_{\alpha \beta}^\pm=\half \left(\epsilon_{\alpha\beta} \pm {\kappa_{\alpha\beta} \over \sqrt{-\kappa^2}} \right),
\qquad
\pi_{\dot \alpha \dot \beta}^\pm 
=\half \left(\epsilon_{\dot \alpha \dot \beta} \pm {\kappa_{\dot \alpha\dot \beta} \over \sqrt{-\kappa^2}} \right).
\eeq
Because of the introduction of the above projectors, the $O(b^2)$ terms vanish. 
Then one can note that with the ansatz for $W$, \eqref{masterfull2} is trivially satisfied. 
At first order in $b$, the second equation of \eqref{wigner} reads
\beq
\frac{\partial}{\partial a^\alpha}\Omega=-\half \mathcal{Q} \sigma_\alpha^\pm\,,
\label{lasteq1lin}
\eeq
and is solved by
\beq
\Omega(a)=f_0+O(b^2)\,,
\eeq
where we use the fact that $\kappa_{\alpha \beta}/r$ is a constant. 
As argued in~\cite{DV}, the $O(b^2)$ terms vanish, and the first 
equation in the last line of \eqref{wigner} then imposes
\beq
df_0=0\,,
\qquad
\label{eqf0}
\eeq
and we take $f_0=0$.

\subsubsection{The Solution}
We can now give the solution to the full non-linear equations of motion
\begin{align}
S_\alpha&=z_\alpha+F_K {e^{i \theta} b\over  r} 
\pi_\alpha^{+\beta}a_\beta
\int_0^1 dt \exp\left(-\frac{it \kappa_{\alpha\beta}}{2  r^2}a^\alpha a^\beta\right)
\label{sol1} \\
\bar{S}_{\dot \alpha}
&=\bar{z}_{\dot \alpha}+F_K {e^{-i\theta} b \over r} 
\pi_{\dot \alpha}^{+\dot \beta}\bar{a}_{\dot \beta}
\int_0^1 dt \exp \left(-{it \kappa_{\dot \alpha \dot \beta} \over 2  r^2}\bar{a}^{\dot \alpha} \bar a^ {\dot \beta} \right) \label{sol2} \\
B&={4b \over  r} \exp \left[-{i \over 2 r^2} 
\left(\kappa_{\alpha\beta}y^\alpha y^\beta
+\kappa_{\dot \alpha \dot \beta}\bar{y}^{\dot \alpha} \bar{y}^{\dot \beta}
+2 i \kappa_{\alpha \gamma}v^\gamma_{\ \dot \beta} y^\alpha \bar{y}^{\dot \beta} \right) \right]
\label{sol3} \\
W&=W_0=-{1 \over 4} \left(\omega_{\alpha \beta} y^\alpha y^\beta 
+\omega_{\dot \alpha \dot \beta} \bar{y}^{\dot \alpha} \bar{y}^{\dot \beta} 
-\sqrt{2} h_{\alpha \dot \beta} y^\alpha \bar{y}^{\dot \beta} \right).
\label{sol4}
\end{align} 
It is then straightforward to show that the solution obtained satisfies the bosonic 
non-minimal reality projection which is obtained from \eqref{susygenproj} by 
removing the $\Gamma$. The bosonic symmetries of the solution are discussed 
in detail in~\cite{DV}. There it is shown that the higher-spin symmetries are broken 
to a subalgebra, whose unique finite dimensional subalgebra is $su(2) \oplus gl(1)$, 
thus hinting at yet another similarity with a black hole solution.

As advertised in the main text, the master-field $W$ \eqref{sol4} is equal to the 
vacuum value $W_0$, and is thus much simpler than the expression given in \cite{DV}. 
This can be traced back to our choice of Killing vector, for which the quantity 
$\kappa_{\alpha \beta}/r$ is a space-time independent constant.

It should also be noted that this solution generalizes slightly the one presented 
in~\cite{DV} as it accommodates for an arbitrary parity breaking phase $\theta_0$. As previously mentioned, the DV solution was generalized in \cite{families}, by working in a different gauge and using an infinite family of projectors similar to \eqref{ProjectorsPi}. Furthermore, the authors of \cite{families} considered an arbitrary real even function for the interaction ambiguity $\theta(X)=\theta_0+\theta_2 X^2+\dots$, instead of just a phase. We show here that this can also be done in the gauge of~\cite{DV}. 
Indeed, using
\beq
B\star v=b F_K\star\delta(y)\star v=b F_K\star\delta(z)=b F_k\cK
\eeq
and since $\mathcal K$ squares to one under the induced star-product \eqref{induced} 
and $F_K$ is a projector we obtain
\bal
f(B \star v)&=1+B \star v \star \exp_\star [i \theta(B \star v )]
=1+B \star v \star \exp_\star [i \theta(bF_K\cK)]
\\&
=1+B \star v \star F_K\exp [i \theta(b)]
=1+B \star v \exp [i \theta(b)]\\
\eal
This means that this particular solution also solves the equations of motion in the 
general case after a redefinition of $\theta_0$. 

\section{Killing Spinors in Global $AdS_4$}
\label{sec:KSpinor}

\subsection{$Sp(4)$ Matrices and Notations}
Here we will try to make clear a few things about notations. We define an 
$Sp(4)$ matrix the following way
\beq
\label{sp4matrix}
M_{AB} = \begin{pmatrix} M_{\alpha \beta} & M_{\alpha \dot \beta} 
\\ M_{\dot\alpha \beta} & M_{\dot \alpha \dot \beta} \end{pmatrix}.
\eeq
Then whenever we use it in expressions involving the star-product and the $Y,Z$, 
we will manipulate the quantity $M_{AB}Y^AY^B$ with 
$Y^A= (y^\alpha, \bar{y}^{\dot \alpha})$. As an example, the 
covariant constancy condition for such a matrix reads $D_0( M_{AB}Y^AY^B)=0$.

Now we define its action on $Y^A$ (or on a spinor 
$\xi^A = (\xi^{\alpha},i \bar{\chi}^{\dot \alpha})$) as $M_A{}^BY_B$. However, it will 
prove to be easier to manipulate $M_{AB}$ as a usual matrix when acting on spinors. 
Then one has to be careful how to go from an expression involving the star-product 
to a matrix expression. To see how this works, we will work with the example of 
imposing that a spinor is covariantly constant. We define 
$\epsilon = \xi_\alpha y^\alpha + i \bar{\chi}_{\dot \alpha} \bar{y}^{\dot \alpha}$. 
Note here the $i$ in the definition. The equation we want to consider is thus 
$D_0 \epsilon=0$. This reads in components
\bal
d \xi_{\alpha} + \omega_{\alpha}{}^{\beta} \xi_\beta 
+ {i \over \sqrt{2}} h_{\alpha \dot \beta} \bar{\chi}^{\dot \beta} &=0 \, , \\
d \bar \chi ^{\dot \alpha} - \omega^{\dot \alpha}{}_{\dot \beta} \bar{\chi}^{\dot \beta} 
+ {i \over \sqrt{2}} h^{\dot \alpha \beta} \xi_{\beta} &=0.
\eal
We then represent the spinor $\epsilon$ as a column vector that we denote 
$\tilde \epsilon$ as follows
\beq
\label{epsilonvector}
\tilde \epsilon = \begin{pmatrix} \xi_\alpha \\ \bar \chi ^{ \dot \alpha} \end{pmatrix}
\eeq
where one should note the absence of $i$, and the position of the indices. Using 
this notation, the Killing spinor equation reads:
\beq
\begin{pmatrix} 1 & 0 \\ 0 & i \end{pmatrix}
\left(d - {i \over 2} \omega_{ab} \gamma^{ab} + {i\over \sqrt{2}} h_a \gamma^a \right) \tilde \epsilon =0.
\eeq

Now we will consider the case of a general $Sp(4)$ matrix $M_{AB}$ again, and 
see how it acts on the column vector $\tilde \epsilon$ \eqref{epsilonvector}. We have
\beq
\xi_AY^A 
\rightarrow M_A{}^B \xi_B Y^A 
= \left( M_{\alpha}{}^{\beta} \xi_{\beta} 
+ M_{\alpha}{}^{\dot \beta} i \bar{\chi}_{\dot \beta} \right) y^\alpha 
+ \left( M_{\dot \alpha}{}^{\dot \beta} i \bar{\chi}_{\dot \beta} 
+ M_{\dot \alpha}{}^{\beta} \xi_{\beta} \right) \bar{y}^{\dot \alpha}
\eeq
Now if we raise and lower indices accordingly to see what this gives on a spinor in the 
notation of \eqref{epsilonvector}, we obtain
\beq
\begin{pmatrix} \xi_\alpha \\ \bar \chi ^{ \dot \alpha} \end{pmatrix}
\ \to\ 
\tilde M \begin{pmatrix} \xi_\alpha \\ \bar \chi ^{ \dot \alpha} \end{pmatrix}
=\begin{pmatrix} 1 & 0 \\ 0 & i \end{pmatrix} 
\begin{pmatrix} M_{\alpha}{}^{\beta} \xi_{\beta} - M_{\alpha \dot \beta} i \bar{\chi}^{\dot \beta} \\
- M^{\dot \alpha}{}_{\dot \beta} i \bar{\chi}^{\dot \beta} + M^{\dot \alpha \beta} \xi_{\beta} \end{pmatrix} 
=\begin{pmatrix} M_{\alpha}{}^{\beta} & - i M_{\alpha \dot \beta} \\
-i M^{\dot \alpha \beta} & - M^{\dot \alpha}{}_{\dot \beta} \end{pmatrix} \begin{pmatrix} \xi_\beta \\ \bar{\chi}^{\dot \beta}\end{pmatrix}
\eeq
So we learn that the matrix
\beq
\label{spinormatrix}
\tilde M = \begin{pmatrix} M_{\alpha}{}^{\beta} & - i M_{\alpha \dot \beta} \\
-i M^{\dot \alpha \beta} & - M^{\dot \alpha}{}_{\dot \beta} \end{pmatrix}
\eeq
acts on $\tilde \epsilon$ and gives back a spinor of the same form.

\subsection{The Killing Spinors of Global $AdS_4$}
\label{sec:KillingSpinorGlobal}
The $AdS_4$ Killing spinor equation is given by
\beq
\left(d - {i \over 2} \omega_{ab} \gamma^{ab}+ {i \over \sqrt{2}} h_a \gamma^a \right)\epsilon =0 \, , 
\eeq
where $\epsilon$ is a $4$-component spinor. We obtain
\bal
\partial_t \epsilon 
&= i \left( r \gamma^{01} - {1 \over \sqrt 2} \sqrt{1+r^2} \gamma^0 \right) \epsilon 
= {i \over \sqrt{2}} \gamma^0 \left( i \sqrt{2} r \gamma^1 - \sqrt{1 + r^2} \right) \epsilon \, , \\
\partial_r \epsilon 
&= { i \over \sqrt 2 \sqrt{1 + r^2}} \gamma^1 \epsilon \, , \\
\partial_\theta \epsilon 
&= i \left( \sqrt{1+r^2} \gamma^{12} + {1 \over \sqrt 2} r \gamma^2 \right) \epsilon \, , \\
\partial_\varphi \epsilon 
& = i \left( \sqrt{1 + r^2} \sin \theta \gamma^{13} + \cos \theta \gamma^{23} 
+ {1 \over \sqrt 2 } r \sin \theta \gamma^3 \right) \epsilon \,.
\eal

The solution to these equations is given by
\beq
\label{KillingSpinors}
\epsilon=\Omega\epsilon_0\,,\qquad
\Omega=e^{ { i \rho \over \sqrt 2} \gamma^1} e^{ - { i t \over \sqrt 2} \gamma^0} 
e^{ i \theta \gamma^{12}} e^{{ i \varphi} \gamma^{23}}\, ,
\eeq
with $\sinh \rho= r$ and $\epsilon_0$ is an arbitrary constant spinor.

An important point is that if 
$\epsilon = \xi_{\alpha}y^\alpha + i \bar{\chi}_{\dot \alpha} \bar{y}^{\dot \alpha}$ 
is a Killing spinor, then so is 
$\epsilon' = \chi_{\alpha}y^\alpha + i \bar{\xi}_{\dot \alpha} \bar{y}^{\dot \alpha}$, 
where we define $\bar{\chi}_{\dot \alpha} = \left( \chi_{\alpha} \right)^\dagger$ 
and $\bar{\xi}_{\dot \alpha} = \left( \xi_{\alpha} \right)^\dagger$. 
With this in mind, a convenient choice of constant spinors leads to the four Killing spinors $\epsilon_I$
\beq
\label{killingspinorbasis}
\epsilon_1 = {\Omega \over \sqrt 2} \begin{pmatrix} 1 \\ 0 \\ 1 \\ 0 \end{pmatrix} \, , 
\quad 
\epsilon_2 = {\Omega \over \sqrt 2} \begin{pmatrix} 0 \\ 1 \\ 0 \\ 1 \end{pmatrix} \, , 
\quad 
\epsilon_{\bar 2} = {\Omega \over \sqrt 2} \begin{pmatrix} -1 \\ 0 \\ 1 \\ 0 \end{pmatrix} \, , 
\quad 
\epsilon_{\bar 1} = {\Omega \over \sqrt 2}\begin{pmatrix} 0 \\ 1 \\ 0 \\ -1 \end{pmatrix} \,.
\eeq
If we write each spinor as in \eqn{epsilonvector}
\beq
\epsilon_i \equiv \begin{pmatrix} \xi_{i , \alpha} \\ \bar \chi_i^{\dot \alpha} \end{pmatrix},
\eeq
then we obtain
\beq
\label{conjugationprop}
\begin{pmatrix} \chi_{1 , \alpha} \\ \bar \xi_1^{\dot \alpha} \end{pmatrix} 
= \begin{pmatrix} \xi_{\bar 1 , \alpha} \\ \bar \chi_{\bar 1}^{\dot \alpha} \end{pmatrix} \, , 
\qquad 
\begin{pmatrix} \chi_{2 , \alpha} \\ \bar \xi_2^{\dot \alpha} \end{pmatrix} 
= \begin{pmatrix} \xi_{\bar 2 , \alpha} \\ \bar \chi_{\bar 2}^{\dot \alpha} \end{pmatrix}.
\eeq

Furthermore, we have 
\begin{gather}
\label{norm}
\bar \epsilon_1 \epsilon_1 = \bar \epsilon_2 \epsilon_2 = - \bar \epsilon_{\bar 2} \epsilon_{\bar 2} = - \bar{\epsilon}_{\bar 1} \epsilon_{\bar 1} =1.
\\
\label{ortho}
\bar \epsilon_i \epsilon_j = 0 \, , \quad i \neq j.
\end{gather}
where we have defined $\bar \epsilon_i = \epsilon_i^\dagger \sqrt{2} \gamma^0$.

\subsection{Killing Vectors of $AdS_4$}

In this section we will find explicitly the Killing vectors of global $AdS_4$ and show that they 
can be put into two distinct categories. The splitting corresponds to the conformal and 
non-conformal isometries of the boundary $R \times S^2$.

First recall that $AdS_4$ can be seen as the embedding in $\mathbb{R}^{2,3}$ of the 
following hyperboloid
\beq
X_0^2 + X_4^2 - \sum_{i=1}^3 X_i^2= 1
\eeq
From this it is easy to see that the isometries of $AdS_4$ are generated by the following Killing vectors
\beq
L_{ab} = X_a {\partial \over \partial X^b} - X_b {\partial \over \partial X^a}
\eeq
If we want to write them using the coordinates of global $AdS_4$ where the metric takes the form
\beq
ds^2 =  \left[ \cosh^2\rho\, dt^2 
- d\rho^2 - \sinh^2 \rho \left( d \theta^2 + \sin^2\theta\, d \varphi^2 \right) \right] \, , 
\eeq
We need to make the following change of variables
\beq
X_0 + i X_4 =  e^{it} \cosh\rho \, , 
\qquad 
X_i =  n_i \sinh\rho
\eeq
with $n_i = (\cos\theta, \sin\theta \cos\varphi, \sin\theta \sin\varphi)$.
We obtain a first set of four Killing vectors which do not depend on $\rho$ nor $t$
\beq
\label{adskillingvectors}
L_{04} = {\partial_t} \, , 
\qquad 
L_{32} = {\partial_\varphi} \, , 
\qquad
L_{21} + i L_{31} = e^{i \varphi} \left( {\partial_\theta} +i \cot\theta \,\partial_\varphi \right).
\eeq
These are also Killing vectors of the boundary $R \times S^2$.
The other six Killing vectors of $AdS_4$ are
\bal
\label{adsconformalkilling}
L_{01} + i L_{14} 
&= e^{it} \left( i \cos\theta \tanh\rho \,\partial_t
+ \cos\theta \,\partial_\rho -\coth\rho \sin\theta \,\partial_\theta \right) \, , 
\\
L_{02} + i L_{24} 
&= e^{it} \Big( i \cos\varphi \sin\theta \tanh\rho \,\partial_t 
+ \cos\varphi \sin\theta \,\partial_\rho  \\
& \qquad + \cos\theta \cos\varphi \coth\rho \,\partial_\theta 
- {\coth\rho\over \sin\theta} \sin\varphi \,\partial_\varphi \Big)\, , \\
L_{03} + i L_{34} 
&= e^{it} \Big( i \sin\theta \sin\varphi \tanh \rho \,\partial_t 
+ \sin\theta\sin\varphi \,\partial_\rho  \\
&\qquad  + \cos\theta \coth\rho \sin\varphi \,\partial_\theta 
+ {\cos\varphi \coth\rho \over \sin\theta} \,\partial_\varphi \Big) \, ,
\eal
and in the $\rho \rightarrow \infty$ become the \emph{conformal} Killing vectors of the boundary manifold.

\subsection{Killing Vectors from Killing Spinors}
It is well known that Killing vectors can be obtained as the bilinears 
$\bar \epsilon' \gamma^\mu \epsilon$ where $\epsilon, \epsilon'$ are two independent Killing spinors and 
$\bar \epsilon = \sqrt{2} \epsilon^\dagger \gamma^0$. We now give explicit expressions for the Killing vectors 
\eqref{adskillingvectors} in terms of the Killing spinors $\epsilon_1,\epsilon_2$ defined in \eqref{killingspinorbasis}
\bal
L_{04}^\mu &= {1 \over \sqrt{2} } \bar \epsilon_1 \gamma^\mu \epsilon_1 +{1 \over \sqrt{2} } \bar \epsilon_2 \gamma^\mu \epsilon_2 \, , 
\qquad 
L_{32}^\mu = - {1 \over \sqrt{2} } \bar \epsilon_1 \gamma^\mu \epsilon_2 - {1 \over \sqrt{2} }\bar \epsilon_2 \gamma^\mu \epsilon_1 \, , \\
 L_{31}^\mu &= {i \over \sqrt 2} \bar \epsilon_1 \gamma^\mu \epsilon_2 - {i \over \sqrt 2} \bar \epsilon_2 \gamma^\mu \epsilon_1 \, , 
 \qquad 
 L_{21}^\mu = - {1 \over \sqrt 2} \bar \epsilon_1 \gamma^\mu \epsilon_1 + {1 \over \sqrt 2} \bar \epsilon_2 \gamma^\mu \epsilon_2 \, , \\
\eal

\subsection{The DV Killing Matrix From Killing Spinors}
\label{sec:DV-Killing}
A central ingredient in \cite{DV} is an $Sp(4)$ matrix which is 
$AdS$ covariantly constant and which squares to minus one, {\em i.e.}, from 
which we can construct a projector. The projector is then a key element in 
simplifying the equations of motion, as well as in proving that the solution is BPS.

Using the notations of the previous subsection, $V_\mu$ of \eqn{Vdt} is
\beq
V_{\gamma \dot \delta} = V^a \left(\sigma_a\right)_{\gamma \dot \delta} 
=\left(\bar \epsilon_1 \gamma^a \epsilon_1 
+ \bar \epsilon_2 \gamma^a \epsilon_2 \right) \left(\sigma_a\right)_{\gamma \dot \delta}
\eeq
Noting that:
\beq
\bar \epsilon_i \gamma^a \epsilon_i = \begin{pmatrix} \chi_i^\alpha 
& \bar{\xi}_{i, \dot \alpha} \end{pmatrix} \begin{pmatrix} 0
& \left( \sigma ^a \right)_{\alpha \dot \beta} \\ 
\left( \bar \sigma ^a \right)^{\dot \alpha \beta}
&0 \end{pmatrix} 
\begin{pmatrix} \xi_{i,\beta} \\ \bar{\chi}_i^{\dot \beta} \end{pmatrix} 
= \chi_i^\alpha \left( \sigma ^a \right)_{\alpha \dot \beta} \bar{\chi}_i^{\dot \beta} 
+ \bar{\xi}_{i, \dot \alpha} \left( \bar \sigma ^a \right)^{\dot \alpha \beta}\xi_{i,\beta}.
\eeq
So that
\beq
V_{\gamma \dot \delta} 
= \sum_{i=1,2} \left( \chi_i^\alpha \left( \sigma ^a \right)_{\alpha \dot \beta} \bar{\chi}_i^{\dot \beta} 
+ \bar{\xi}_{i, \dot \alpha} \left( \bar \sigma ^a \right)^{\dot \alpha \beta}\xi_{i,\beta} \right) 
\left(\sigma_a\right)_{\gamma \dot \delta}
= \sum_{i=1,2} \left( \chi_{i,\gamma} \bar{\chi}_{i,\dot \delta} 
+ \xi_{i,\gamma} \bar{\xi}_{i, \dot \delta} \right) 
\eeq
Now if we define
\beq
\label{semiprojector}
\varkappa_{i\bar i,AB} \equiv 2 \epsilon_{i,A} \epsilon_{\bar i, B} 
=\begin{pmatrix} \xi_{i,\alpha} \chi_{i,\beta} + \chi_{i,\alpha} \xi_{i,\beta} 
& \xi_{i, \alpha} i \bar{\xi}_{i,\dot \beta} + \chi_{i, \alpha} i \bar{\chi}_{i,\dot \beta} \\ \chi_{i,\alpha} i \bar{\chi}_{i,\dot \beta} + \xi_{i,\alpha} i \bar{\xi}_{i,\dot \beta} & i \bar{\chi}_{i,\dot \alpha} i \bar{\xi}_{i,\dot \beta} +i \bar{\xi}_{i,\dot \alpha} i \bar{\chi}_{i,\dot \beta} \end{pmatrix} \, , 
\eeq
then we see that the diagonal entries are reminiscent of the time-like Killing 
vectors expressed in terms of the components of the basis of Killing spinors. 
The corresponding operators which acts on the Killing spinors basis 
\eqref{killingspinorbasis} is
\beq
P_i \equiv \epsilon_i \bar \epsilon_i + \epsilon_{\bar i} \bar{\epsilon}_{\bar i}.
\eeq
However one should note that for this particular Killing vector, there is a sum over 
$i=1,2$. This amounts to taking $P_1+P_2$. Indeed, this yields a matrix which is 
by construction covariantly constant, and which also squares to the identity, since 
$\left(P_1 + P_2\right)^2 = \epsilon_1 \bar{\epsilon}_1 + \epsilon_2 \bar{\epsilon}_2 
- \epsilon_{\bar 1} \bar{\epsilon}_{\bar 1} - \epsilon_{\bar 2} \bar{\epsilon}_{\bar 2} =1$, 
when acting on the space spanned by \eqref{killingspinorbasis}. In terms of $Sp(4)$ 
matrices, this is $\varkappa_{1\bar1, AB}+\varkappa_{2 \bar 2, AB}$. 
If we look at the explicit content 
of this matrix, it is straightforward to relate it to the Killing matrix $K_{AB}$ 
defined in \eqref{KillingDV}
\beq
\varkappa_{1\bar1, AB}+\varkappa_{2 \bar 2, AB} = i K_{AB} \, ,
\eeq
just as expected. This allows us to see clearly how $K_{AB}$, as well as the 
projector $\Pi_{\pm,AB}$ defined in \eqref{ProjectorsPi} act on our basis of 
Killing spinors \eqref{killingspinorbasis}. We see using \eqref{conjugationprop} 
and \eqref{norm} that $\Pi_+$ projects on the space spanned by 
$\epsilon_1,\epsilon_2$, while $\Pi_-$ projects on the space spanned 
by $\epsilon_{\bar 1}, \epsilon_{\bar 2}$.

\section{Representations of the Clifford Algebra}
\label{sec:Matrices}

We choose the following matrix representation for the $\vartheta^i$ in the $n=4$ case
\bal
\vartheta^1 &= \frac{1+i}{\sqrt 2}\begin{pmatrix}
0 & 0 &\ 0 & -1\\ 
0 & 0 &\ 1& 0 \\
0 & -i&\ 0 & 0\\
 i& 0 &\ 0 & 0 \\
\end{pmatrix} \, , 
&\quad 
\vartheta^2 &=\frac{1+i}{\sqrt 2}\begin{pmatrix}
0 & 0 &\ 0 &\ i \\ 
0 & 0 &\ i &\  0 \\
0 & -1&\ 0 &\ 0\\
 -1& 0 &\ 0 &\ 0 \\
\end{pmatrix} \, , \\
\vartheta^3 &=\frac{1+i}{\sqrt 2}\begin{pmatrix}
0 & 0 &\ i &\ 0 \\ 
0 & 0 &\ 0 &-i \\
-1& 0 &\ 0 &\ 0\\
0 & 1&\ 0 &\ 0 \\
\end{pmatrix} \, , 
&\quad 
\vartheta^4 &=\frac{1+i}{\sqrt 2}\begin{pmatrix}
0 & 0 & \ 1&\ 0 \\ 
0 & 0 &\ 0 &\ 1\\
-i & 0 &\ 0 &\ 0\\
0 & -i &\ 0 &\ 0 \\
\end{pmatrix} \, , \\
\eal

\end{document}